\documentclass[aps,prb,twocolumn,amsmath,amssymb,superscriptaddress,floatfix,scrartcl]{revtex4-1}

\usepackage{amssymb}
\usepackage{color}
\usepackage{graphicx}
\usepackage{srcltx}
\usepackage{mathrsfs}
\usepackage[unicode=true,pdfusetitle,
 bookmarks=false,colorlinks=true,citecolor=blue,urlcolor=blue,linkcolor=red]{hyperref}

\begin{document}

\title{
Wilson operator algebras and ground states of coupled BF theories
   }

\author{Apoorv Tiwari}
\email{t.apoorv@gmail.com}
\affiliation{
Institute for Condensed Matter Theory and
Department of Physics, University of Illinois
at Urbana-Champaign,
1110 West Green St, Urbana IL 61801
            }
\thanks{The first two authors contributed equally to the work.}

\author{Xiao Chen}
\email{chenxiao.phy@gmail.com}
\affiliation{
Institute for Condensed Matter Theory and
Department of Physics, University of Illinois
at Urbana-Champaign,
1110 West Green St, Urbana IL 61801
            }

\author{Shinsei Ryu}
\email{ryuu@illinois.edu}
\affiliation{
Institute for Condensed Matter Theory and
Department of Physics, University of Illinois
at Urbana-Champaign,
1110 West Green St, Urbana IL 61801
            }

\date{\today}

\begin{abstract}
The multi-flavor $BF$ theories in (3+1) dimensions with cubic or quartic coupling
are the simplest topological quantum field theories that can describe fractional braiding statistics
between loop-like topological excitations (three-loop or four-loop braiding statistics).
In this paper, by canonically quantizing these theories, we study
the algebra of Wilson loop and Wilson surface operators,
and multiplets of ground states on three torus.
In particular, by quantizing these coupled $BF$ theories on the three-torus,
we explicitly calculate
the $\mathcal{S}$- and $\mathcal{T}$-matrices, which encode fractional braiding statistics and
topological spin of loop-like excitations, respectively.
In the coupled $BF$ theories with cubic and quartic coupling, the Hopf link and Borromean ring of loop excitations,
together with point-like excitations, form composite particles.
\end{abstract}


\maketitle

\tableofcontents

\section{Introduction}

For more than three decades exotic quantum phases of matter have been extensively studied in condensed matter physics.
In particular gapped systems with non-trivial topological order have been of much interest.
\cite{Wen04quantumfield}
Topologically ordered phases have properties such as fractional statistics, long-range entanglement, ground-state degeneracy on manifolds with non-trivial topology,
and symmetry fractionalization, etc.
\cite{wen1990ground,nayak2008non,levin2006detecting,kitaev2006topological,dong2008topological,fendley2007topological}
Canonical examples are fractional quantum Hall states in 2+1 dimensions,
which have been observed experimentally.
%These phases of matter have been observed in 2+1 dimensions and canonical examples are the fractional quantum Hall state.

At long wavelengths,
topologically ordered phases of matter can be described by topological quantum field theories (TQFTs),
for which all correlation functions are topological,
i.e., metric independent.
For example,
many fractional quantum Hall states,
as well as simple lattice models such as the Kitaev toric code model,
\cite{kitaev2003fault,levin2005string,walker20123+}
can be described by the Chern-Simons topological quantum field theories.
For these examples,
fractional braiding statistics between quasiparticles is described in terms of Wilson lines (loops) in the TQFTs,
i.e.,
by the correlation functions of Wilson loops forming a Hopf link in the $2+1$-dimensional spacetime.
\cite{Witten1989}
%The correlation functions of these Wilson loops can be understood as forming Hopf links in the $2+1$-dimensional spacetime manifold.

%In $2+1$ dimensions, topologically ordered phases exhibit fractional braiding statistics between particles.
%world-lines.
%Some simple examples include the toric code model and the double semion model.
%These topological phases can be described by the Chern-Simons field theory, where the observables are Wilson lines (loops).
%The correlation functions of these Wilson loops can be understood as forming Hopf links in the $2+1$-dimensional spacetime manifold.

%\textcolor{red}{Mention classic papers by Witten, quantization of CS, bulk boundary?}

The idea of fractional braiding statistics can be generalized to $3+1$ dimensions.
Since particles cannot braid in three spatial dimensions or equivalently, their world-lines cannot link in $3+1$-dimensions,
the simplest kind of braiding is between point-link and loop-like excitations, which can have non-trivial fractional braiding statistics.
This is described by the $BF$ topological field theory and has been studied quite well.
\cite{Horowitz1989, Horowitz1990, Balachandran1993, BergeronSemenoffSzabo1995, Szabo1998}
Topological phases in $3+1$-dimensions, however, have richer possibilities in terms of the kind of braiding processes that can exist.
\cite{WangLevin2015, WangWen2015, JiangMesarosRan2014, WangLevin2014,wang2016quantum, wan2015twisted}

In this work we explore a subset of such processes by using  (3+1)-dimensional TQFTs.
%\textcolor{red}{$3+1$-dimensional Dijkgraaf Witten theories.}
In particular, we study TQFTs which can be thought of as extensions of the ordinary $BF$ theory.
We mainly study two kinds of extensions:
%The first is the usual $BF$ theory which describes braiding between particles and loops.
The first is the $BF$ theory with a cubic deformation.
More precisely, we consider multiple (two or three) copies of the $BF$ theory coupled together via a cubic term.
These theories realize non-trivial statistics between three loop excitations whose spacetime world surfaces are linked together,
i.e., the so-called the three-loop braiding statistics.
The second is four (or more) copies of $BF$ theories coupled via quartic terms.
These field theories describe four-loop braiding statistics.
Similar TQFTs with cubic and quartic coupling terms have been discussed recently in the literature.
\cite{kapustin2014, wang2016quantum}
The coupled $BF$ theories with cubic or quartic coupling can also be obtained by functionally bosonizing (or gauging) bosonic symmetry protected phase (SPT) described in Ref.\ \onlinecite{YeGu2015,JWWang2015}.

In addition to these TQFTs with a cubic or quartic coupling,
we will also discuss avatars of these coupled topological field theories
which are quadratic but with modified coupling to external currents.
We will quantize these quadratic theories on the spatial three torus and discuss the algebra of Wilson operators over there.

%It is important to note that each of these braiding processes are fundamental in the sense that a four loop braiding cannot be decomposed into a three-loop braiding process and so on.
%An enumeration of the kinds of fundamental braiding processes and the corresponding quantum field theories that describe such processes remains an open problem.

A salient feature of topological field theories is bulk boundary correspondence
wherein
ground states in the bulk Hilbert space
are in one-to-one correspondence
with
twisted partition functions defined for the boundary field theory.
In our previous work,
\cite{chen2015bulk}
we studied the two-copies of $BF$ theories coupled by a cubic term,
but focused on the gapless surface theory and the boundary-bulk correspondence:
We quantized the surface theory and explicitly calculated the partition functions under various twisted boundary conditions.
In addition, by performing large diffeomorphism transformations or modular transformations on the twisted partition functions,
we extracted the bulk braiding data directly from the gapless surface theory.
(As a related work, see Ref.\ \onlinecite{2015arXiv151209111W} for the bulk-boundary correspondence for gapped topologically ordered surface states.)
In this work we study such TQFTs describing three-loop and four-loop braiding in more detail.
In particular, we will study various ``bulk'' properties of these TQFTs, and hence provide
a complementary perspective to our previous work.

\subsection{Summary and outline}

The summary of our main results, as well as the outline of the paper, is given as follows.

Section \ref{Three-loop braiding theory} is devoted to the coupled $BF$ theories realizing
non-trivial three-loop braiding statistics.
In Sec.\ \ref{The cubic theories} and Sec.\ \ref{The quadratic theory},
we introduce these coupled $BF$ theories, and give an overview of their basic properties.
In particular,
at the classical level, one can read off from the equations of motion that Hopf links
%and Borromean rings
play particle-like roles in these two theories.
%i.e.,  they braid with loops to give three loop and four loop braiding, respectively.
This braiding structure is encoded in the algebra of the dynamical gauge fields in these theories.

In the following Sections
\ref{Three-loop braiding statistics},
\ref{Quantization on a closed spatial manifold}, and \ref{Wave function in terms of Wilson operators},
we quantized the quadratic $BF$ theories introduced in Sec.\ \ref{The quadratic theory},
which differ from the ordinary $BF$ theory
due to their modified coupling to the quasi-vortex current.
%Instead of directly quantizing, the cubic and quartic TQFTs,
%we propose
The quadratic theory has the same equations of motion as the cubic theories.
%alternative quadratic theories which we claim our equivalent to the original cubic and quartic theories.
%The quadratic theories have the same equations of motion as the respective original theories.
Moreover, the Wilson operator algebra of the quadratic theories encodes the three-loop braiding statistics.
More specifically the commutator, and triple commutator between the respective Wilson operators
are relevant to the respective particle-loop, and the three-loop braiding phases
(Sec. \ref{Three-loop braiding statistics}).

Further we quantize the quadratic three-loop braiding field theory on a spatial three torus
in Sections \ref{Quantization on a closed spatial manifold} and \ref{Wave function in terms of Wilson operators}.
We construct the multiplet of ground states
of the two (or three) copies of the $BF$ theories at level $\mathrm{K}$ put on spatial three torus $T^3$,
by directly constructing representations of the Wilson operator algebra.
The ground state degeneracy is $\mathrm{K}^2$ (or $\mathrm{K}^3$).
In Appendix \ref{Ground state wave functionals by geometric quantization},
an alternative construction of the ground state multiplet by using geometric quantization is given.
Furthermore, by calculating various overlaps between ground states,
we explicitly compute the modular $\mathcal S$ and $\mathcal T$ matrices and extract particle-loop and three-loop braiding phases from them.
These agree with the braiding phases computed in our previous work
from the surface theory,
\cite{chen2015bulk}
as well as
with previous bulk calculations in the literature.
\cite{chen2015bulk, WangLevin2015, WangWen2015, JiangMesarosRan2014,WangLevin2014}
%We perform modular $SL(3,\mathbb Z)$
%{\footnote{The mapping class group of a three-torus $\mathbb T^3$ is $SL(3\mathbb Z)$ which has two generators $\mathcal S$ and $\mathcal T$.}}
%transformations on the ground state multiplet.
%Under modular $\mathcal S$ and $\mathcal T$ transformations,
%the ground state multiplet transforms projectively and the matrix
%of projective phases encodes braiding phases corresponding to particle-loop and three-loop braiding.
%We explicitly compute these $\mathcal S$ and $\mathcal T$ matrices and  extract particle-loop and three-loop braiding phases from them.
%These agree with the braiding phases computed in our previous work directly from the surface theory as well as with previous bulk calculations in the literature \cite{chen2015bulk, WangLevin2015, WangWen2015, JiangMesarosRan2014,WangLevin2014}.

Much of what is discussed in Sec.\ \ref{Three-loop braiding theory}
carries over to Sec.\ \ref{Four-loop braiding theory}, in which we discuss the coupled $BF$ theories realizing
non-trivial four-loop braiding statistics.
In these theories,
the role played by Hopf links in three-loop braiding theories
is played by Borromean rings of loop-like excitations. The role of the triple is replaced by the quadruple commutator of the Wilson operators. This carries information about four loop braiding.

Finally in Sec.\ \ref{Condensation picture},
we propose condensation mechanisms by which topological field theories describing three-loop and four-loop braiding
statistics may arise at long wavelengths.
It is known that the simplest continuum topological field theory in $3+1$ dimensions, i.e.,
the $BF$ theory at level $\mathrm{K}$,
describes the deconfined phase of the $\mathbb Z_{\mathrm{K}}$ gauge theory.
This may arise from a parent (ultraviolet) $U(1)$ gauge theory, if the $U(1)$ gauge symmetry is Higgsed to $\mathbb Z_{\mathrm{K}}$
by the abelian Higgs mechanism.
Alternatively the $BF$ theory may arise as a result of the magnetic condensation via the Julia-Toulouse mechanism.
In Sec.\ \ref{Condensation picture},
we discuss how the coupled $BF$ theories realizing three- or four-loop braiding statistics may arise
from ultraviolet theories by condensation of some sort.
%based on the Julia-Toulouse mechanism by starting with several copies of $U(1)$ gauge theory.
By condensing a composite of electric charge and a Hopf link between $U(1)$ field lines,
it can be shown that the long wavelength effective field theory is a topological field theory that describes three-loop braiding.
Alternately by condensing a composite of electric charge and a Borromean ring between $U(1)$ field lines,
it can be shown that the effective field theory is a topological field theory that describes four-loop braiding.

We conclude in Sec.\ \ref{Conclusion and remarks} with a few words on open issues.

\section{Three-loop braiding theory}
\label{Three-loop braiding theory}

\subsection{The cubic theories}
\label{The cubic theories}

In our previous work,
\cite{chen2015bulk}
we analyzed the coupled $BF$ theory
defined by the following action:
\begin{align}
S&=
\int_{\mathcal M}
\Bigg[\frac{\mathrm{K}}{2\pi}\delta_{IJ}b^{I}\wedge {d}a^{J}
\nonumber \\
&\qquad
+
\frac{ \mathrm{p}_1}{4\pi^2} a^{1}\wedge a^{2}\wedge da^{2}
+
\frac{\mathrm{p}_2}{4\pi^2}  a^{2}\wedge a^{1}\wedge da^{1}
 \nonumber \\
&\qquad
-\delta_{IJ}b^{I}\wedge J^J_{qv}
-\delta_{IJ}a^{I}\wedge J^J_{qp}
\Big],
\label{action 3loop}
\end{align}
where
$a^I$ and $b^I$ are one- and two-form gauge fields, respectively;
$I,J=1,2$;
$\mathcal{M}$ is the (3+1)-dimensional spacetime manifold,
and we will mostly assume $\mathcal{M}=\Sigma\times \mathbb{R}$
where $\Sigma$/$\mathbb{R}$ is a spatial/temporal part of the manifold.
$\mathrm{K}$ and $\mathrm{p}_{1,2}$ are the parameters of the theory;
The ``level'' $\mathrm{K}$ is an integer, whereas
$\mathrm{p}_{1,2}$ are an integer multiple of $\mathrm{K}$
and are given by
\begin{align}
\mathrm{p}_1 = \mathrm{q}_1 \mathrm{K},
\quad
\mathrm{p}_2 = \mathrm{q}_2 \mathrm{K},
\quad
\mathrm{q}_{1,2}=0, \ldots, \mathrm{K}-1.
\end{align}
Finally,
the three-form $J_{qp}$ and two-form $J_{qv}$ represent
quasi-particle and quasi-vortex (loop-like) currents,
which are treated as a non-dynamical background.
For a quasi-particle whose world line is given by $\mathcal{C}\subset \mathcal{M}$,
and for a quasi-vortex whose world surface is given by $\mathcal{S}\subset \mathcal{M}$,
$J_{qp}$ and $J_{qv}$ are given as
\begin{align}
J_{qp} = \delta (\mathcal{C}),
\quad
J_{qv} = \delta(\mathcal{S}),
\end{align}
respectively, where the delta function forms $\delta(\mathcal{C})$ and $\delta(\mathcal{S})$ are defined such that
$
\int_{\mathcal{M}} \delta(\mathcal{C})\wedge A
=\int_{\mathcal{C}} A
$
and
$
\int_{\mathcal{M}} \delta(\mathcal{S})\wedge B
=\int_{\mathcal{S}}B
$
for arbitrary one- and two-form $A$ and $B$, respectively.
(For properties of the delta function forms, see Ref.\ \onlinecite{chen2015bulk}.)

The action \eqref{action 3loop} describes topological gauge theories of various kinds
with gauge group $G=\mathbb Z_{\mathrm{K}}\times \mathbb Z_{\mathrm{K}}$.
Following the seminal work of Dijkgraaf and Witten,
\cite{dijkgraaf1990topological}
we know that topological gauge theories in $d+1$-dimensions with a discrete gauge group $G$
are classified by the group cohomology $H^{d+1}(G,U(1))$.
Since $H^{4}(\mathbb Z_{\mathrm{K}}\times \mathbb Z_{\mathrm{K}},U(1))=\mathbb Z_{\mathrm{K}}\times \mathbb Z_{\mathrm{K}}$,
we expect there are $\mathrm{K}^2$ distinct theories.
Within the coupled $BF$ theory \eqref{action 3loop}, these are parametrized by $\mathrm{p}_{1,2}$
(or equivalently $\mathrm{q}_{1,2}$).

For later use, we record the equations of motion derived from \eqref{action 3loop}:
\begin{align}
\frac{\mathrm{K}}{2\pi}da^{I}&=
J_{qv}^{I},
\nonumber \\
\frac{\mathrm{K}}{2\pi}db^{I}&=-\frac{\mathrm{p}_I}{4\pi^2}a^{\bar{I}}\wedge da^{\bar{I}}
+\frac{\mathrm{p}_{\bar{I}}}{2\pi^2}a^{\bar{I}}\wedge da^I
\nonumber \\
& \quad
-\frac{\mathrm{p}_{\bar{I}}}{4\pi^2}da^{\bar{I}}\wedge a^I
+
J_{qp}^I,
%\nonumber \\
%\frac{\mathrm{K}}{2\pi}db^{2}&=
%-\frac{\mathrm{p}_2}{4\pi^2}a^1\wedge da^1
%+\frac{\mathrm{p}_1}{2\pi^2}a^1\wedge da^2
%\nonumber \\
%& \quad
%-\frac{\mathrm{p}_1}{4\pi^2}da^1\wedge a^2+J_{qp}^2,
\label{eom_cubic}
\end{align}
where we introduced the notation $\bar{1}=2$ and $\bar{2}=1$, and
the repeated capital Roman indices are not summer over here.

In addition to the two flavors of BF theories \eqref{action 3loop},
we will also discuss three flavors of BF theories and couple them by introducing
a cubic term. This leads to the action
\begin{align}
S&=
\int_{\mathcal M}
\Bigg[\frac{\mathrm{K}}{2\pi}\delta_{IJ}b^{I}\wedge {d}a^{J}
+
\mathrm{p} a^{1}\wedge a^{2}\wedge da^{3}
 \nonumber \\
&\qquad
-\delta_{IJ}b^{I}\wedge J^J_{qv}
-\delta_{IJ}a^{I}\wedge J^J_{qp}
\Big],
\label{cubic bf three flavor}
\end{align}
where the flavor indices $I,J$ run over $1,2,3$.
As before, $\mathrm{K}$ and $\mathrm{p}$ are the parameters of the theory.
This three-flavor theory shares similar properties as the two-flavor theory \eqref{action 3loop},
and can be discussed in parallel with the two-flavor theory.
In particular,
both two-flavor and three-flavor theories realize non-trivial three-loop braiding statistics.

%
%The equations of motion are
%\begin{align}
%\frac{K}{2\pi}da^{I}=&\;J_{qv}^{I} \nonumber \\
%\frac{K}{2\pi}db^{1}+\frac{p_1}{4\pi^2}a^2\wedge da^2-\frac{p_2}{2\pi^2}a^2\wedge da^1+\frac{p_2}{4\pi^2}da^2\wedge a^1=&\;J_{qp}^1 \nonumber \\
%\frac{K}{2\pi}db^{2}+\frac{p_2}{4\pi^2}a^1\wedge da^1-\frac{p_1}{2\pi^2}a^1\wedge da^2+\frac{p_1}{4\pi^2}da^1\wedge a^2=&\;J_{qp}^2 \nonumber
%\end{align}
%It is illustrative to rewrite the equations of motion as
%\begin{align}
%da^{I}=&\; \frac{2\pi}{K}J_{qv}^I \nonumber \\
%db^{1}=&\; \frac{2\pi}{K}J_{qp}^1+\frac{2\pi p_1}{K^3}J_{qv}^{2}\wedge d^{-1}J_{qv}^2-\frac{4\pi p_2}{K^3}J_{qv}^1\wedge d^{-1}J_{qv}^{2} \nonumber \\
%&\; -\frac{2\pi p_2}{K^3}J_{qv}^2 \wedge d^{-1}J_{qv}^1 \nonumber \\
%db^{2}=&\; \frac{2\pi}{K}J_{qp}^2+\frac{2\pi p_2}{K^3}J_{qv}^{1}\wedge d^{-1}J_{qv}^1-\frac{4\pi p_1}{K^3}J_{qv}^2\wedge d^{-1}J_{qv}^{1} \nonumber \\
%&\; -\frac{2\pi p_1}{K^3}J_{qv}^1 \wedge d^{-1}J_{qv}^2 \nonumber
%\end{align}
%Here $J_{qv}d^{-1}J_{qv}^{J}$ indicates a Hopf-link between $J_{qv}^{I}$ and $J_{qv}^{J}$. The equations of motions indicate the flux corresponding to 1-form gauge field $a^{I}$ is attached to the quasivortices and flux corresponding to 2-form gauge field $b^{I}$ is attached to composites of quasi-particles and Hopf-linked quasivortices.
%

\subsubsection{Gauge invariance}

Let us now discuss the gauge symmetries of the theory \eqref{action 3loop}.
(We will focus on infinitesimal or small gauge transformations here;
we will discuss large gauge transformations in detail later.)
We first switch off the coupling to currents $J_{qp}$ and  $J_{qv}$.
The action \eqref{action 3loop} is invariant under
\begin{align}
b^{I}&\to b'^I-\frac{\mathrm{p}_{\bar{I}}}{2\pi \mathrm{K}}(a^{\bar{I}}\wedge d\varphi^I
+d\varphi^{\bar{I}}\wedge a^I),
\nonumber \\
%b^{2}&\to b'^2-\frac{\mathrm{p}_1}{2\pi \mathrm{K}}(a^1\wedge d\varphi^2+d\varphi^1\wedge a^2),
%\nonumber \\
a^{I}&\to a'^{I}=a^{I}+d\varphi^{I},
\label{gauge3loop}
\end{align}
where $\varphi^{I}$ is a scalar.
This transformation is a generalization of the usual 1-form gauge symmetry that the ordinary $BF$ theory has.
As in the ordinary $BF$ theory,
the action \eqref{action 3loop} is invariant under an additional 2-form gauge symmetry
\begin{align}
b^{I}\to b^{I}+d\zeta^{I}
\end{align}
where $\zeta^I$ is a one-form.
Formally, these transformations can be read off by identifying the operators that generate the Gauss law constraints.

\begin{figure}[bt]
\centering
\includegraphics[scale=.4]{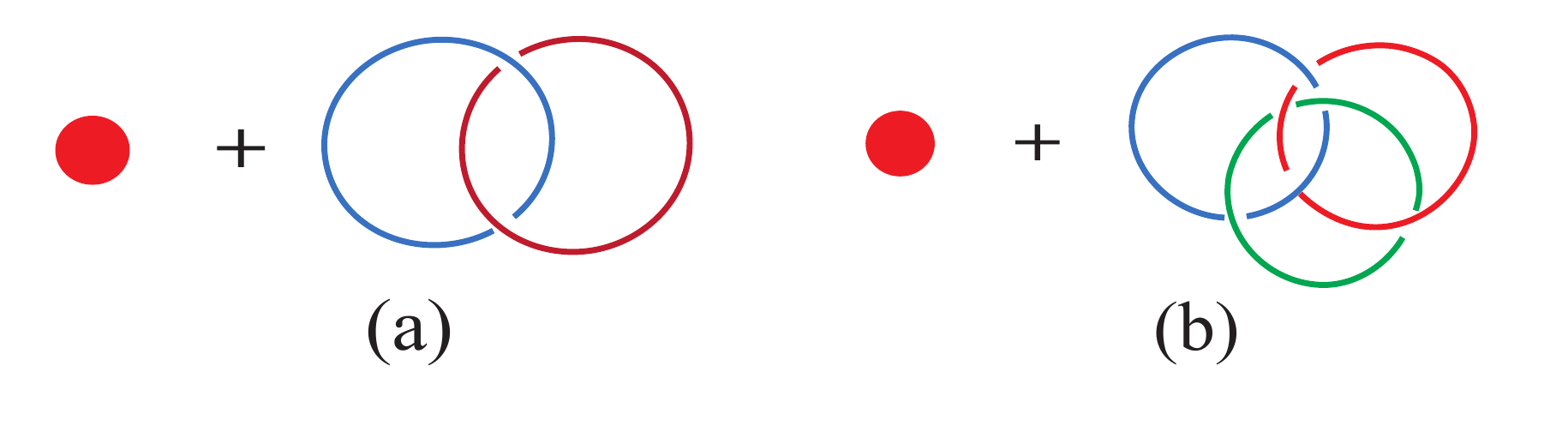}
\\
\caption{
Hopf links (a) and Borromean rings (b) as an effective quasiparticle.
The red dot represents an ``ordinary'' point-like quasiparticle.
%(c-e) Point-like quasiparticles, Hopf links, Borromean rings can braid
%non-trivially with a loop-like excitation (quasivortex) represented by a black solid line.
}
\label{fig:fig0}
\end{figure}

Naively it seems that the coupling to sources in Eq.\ \eqref{action 3loop}
is not gauge invariant.
Upon gauge transformation,
the source terms transform as
\begin{align}
& \delta_{IJ}a^{I}\wedge J_{qp}^{J} + \delta_{IJ}b^{I}\wedge J_{qv}^{J}
\nonumber \\
&\longrightarrow
\delta_{IJ}a^{I}\wedge J_{qp}^{J}+ \delta_{IJ}b^{I}\wedge J_{qv}^{J}
\nonumber \\
&\quad
+d\varphi^{1}\wedge \left[J_{qp}^1+\frac{\mathrm{p}_2}{\mathrm{K}^2}d^{-1}J_{qv}^2\wedge J_{qv}^{1}
-\frac{\mathrm{p}_1}{\mathrm{K}^2}d^{-1}J_{qv}^2\wedge J_{qv}^{2}\right]
\nonumber \\
&\quad
+d\varphi^{2}\wedge \left[J_{qp}^2
+\frac{\mathrm{p}_1}{\mathrm{K}^2}d^{-1}J_{qv}^1\wedge J_{qv}^{2}
-\frac{\mathrm{p}_2}{\mathrm{K}^2}d^{-1}J_{qv}^1\wedge J_{qv}^{1}\right]
\nonumber \\
&\quad
+\delta_{IJ}d\zeta^{I}\wedge J_{qv}^{J},
\end{align}
where we have used the equation of motion
\eqref{eom_cubic}
to write
$a^I = (2\pi/\mathrm{K})(d^{-1} J^I_{qv})$.
Demanding the gauge invariance, we can read off the conservation law of currents,
\begin{align}
&
d\left[J_{qp}^I+\frac{\mathrm{p}_{\bar{I}}}{\mathrm{K}^2}d^{-1}J_{qv}^{\bar{I}}
\wedge J_{qv}^{I}
-\frac{\mathrm{p}_I}{\mathrm{K}^2}d^{-1}J_{qv}^{\bar{I}}\wedge
J_{qv}^{\bar{I}}\right]=0,
%\nonumber \\
%&
%d\left[J_{qp}^2+\frac{\mathrm{p}_1}{\mathrm{K}^2}d^{-1}J_{qv}^1\wedge J_{qv}^{2}
%-\frac{\mathrm{p}_2}{\mathrm{K}^2}d^{-1}J_{qv}^1\wedge J_{qv}^{1}\right]=0,
\nonumber \\
&
dJ_{qv}^{I}= 0.
\label{current conservation}
\end{align}
%In subsequent sections, we will show that [\ref{action 3loop}] describes 3-loop braiding in addition particle-loop braiding.
Here,
for static configuration of currents $J^I_{qv}$,
$d^{-1} J^I_{qv}\wedge J^J_{qv}$,
once integrated over space,
is the Hopf linking number,
\begin{align}
\mbox{Hopf}(J^I_{qv}, J^J_{qv})
=
\int_{\Sigma} (d^{-1} J^I_{qv})\wedge J^J_{qv},
\end{align}
in the spatial manifold $\Sigma$.
Thus, the composite of the particle current and Hopf linking number current is conserved.
This suggests that
the Hopf linking number can be treated effectively as a quasiparticle of some sort
(Fig.\ \ref{fig:fig0}).
%and as we will see momentarily,
%the non-trivial three-loop braiding statistics can be understood as arising from
%exchanging a loop-like excitation and the Hopf linking number.

This point of view also played a crucial role in our previous work, Ref.\ \onlinecite{chen2015bulk}.
Integrating over the equation of motion \eqref{eom_cubic} over the spatial manifold $\Sigma$,
again by using $a^I = (2\pi/\mathrm{K}) (d^{-1}J^I_{qv})$,
we obtain
\begin{align}
\frac{\mathrm{K}}{2\pi} \int_{\Sigma} db^I
&=
-\frac{ \mathrm{p}_I}{\mathrm{K}^2}
\int_{\Sigma} (d^{-1}J^{\bar{I}}_{qv}) \wedge J^{\bar{I}}_{qv}
\nonumber \\
&\quad
+\frac{ \mathrm{p}_{\bar{I}}}{\mathrm{K}^2}
\int_{\Sigma} (d^{-1}J^{\bar{I}}_{qv})\wedge J^I_{qv}
%-\frac{ \mathrm{p}_2}{\mathrm{K}^2} \int_{\Sigma} J^2 \wedge (d^{-1}J^1)
+ \int_{\Sigma} J_{qp}^I,
%\nonumber \\
%\frac{\mathrm{K}}{2\pi} \int_{\Sigma} db^2
%&=
%-\frac{\mathrm{p}_2}{\mathrm{K}^2} \int_{\Sigma} (d^{-1}J^1_{qv})\wedge J^1_{qv}
%\nonumber \\
%&\quad
%+\frac{\mathrm{p}_1}{\mathrm{K}^2} \int_{\Sigma} (d^{-1}J^1_{qv})\wedge J^2_{qv}
%%-\frac{\mathrm{p}_1}{\mathrm{K}^2} \int_{\Sigma} J^1\wedge (d^{-1}J^2)
%+ \int_{\Sigma} J_{qp}^2,
\label{eom integrated}
\end{align}
where note that in the static configurations considered here,
$J_{qv}$ is a delta function two form supporting a spatial loop,
whereas
$J_{qp}$ is a delta function three form supporting a spatial point.
Correspondingly, $d^{-1}J_{qv}$ is a delta function one form supporting
a three dimensional manifold.
The contributions to the flux $\int_{\Sigma}db^I$ coming
from quasivortex loops,
$\int_{\Sigma} (d^{-1}J^I_{qv}) \wedge J^J_{qv}$,
are given in terms of their Hopf linking number.
By using the Stokes theorem,
Eq. \eqref{eom integrated} can be used to link
the twisted partition functions on the boundary and
the quantum numbers in the bulk, and hence to establish the bulk-boundary correspondence.

\subsection{The quadratic theory}
\label{The quadratic theory}

In Ref.\ \onlinecite{chen2015bulk},
an alternative to the cubic theory \eqref{action 3loop},
the quadratic theory,  is proposed:
\begin{align}
	S &=
	\frac{\mathrm{K}}{2\pi}
	\int \delta_{IJ} b^I \wedge da^I
- \int \delta_{IJ} a^I \wedge J^I_{qp}
\nonumber \\
&\quad
-\int \left[ b^1  + \frac{\mathrm{p}_2}{2\pi\mathrm{K}} a^1 \wedge a^2 \right] \wedge J^1_{qv}
\nonumber \\
&\quad
-\int \left[ b^2  + \frac{\mathrm{p}_1}{2\pi\mathrm{K}} a^2 \wedge a^1 \right] \wedge J^2_{qv}.
\label{quadratic 3loopaction}
\end{align}
Comparing the cubic and quadratic theories,
in the cubic theory,
the canonical commutation relations
differ from the ordinary $BF$ theory,
while they remain the same in the quadratic theory.
On the other hand, the set of Wilson loop and surface operators
in the cubic theory is  conventional (i.e., identical to the ordinary $BF$ theory)
while it is modified in the quadratic theory, as seen from the coupling to $J^I_{qv}$
(see below in Eq.\ \eqref{surface op. quadratic theory}).
In spite of these differences,
the algebra of Wilson loop and surface operators of
the two theories appear to be identical.
We will use the cubic and quadratic theories somewhat interchangeably;
When discussing the Wilson operator algebra and ground state wave functions (functionals),
we will use the quadratic theories, while when discussing the condensation picture, we will
use the cubic theory.

\subsubsection{Gauge invariance}

One can derive the infinitesimal gauge transformations from the source-free part of the action \eqref{quadratic 3loopaction}.
Since in this case the theory is identical to the ordinary $BF$ theory,
there are two conserved charges
$(\mathrm{K}/2\pi)db^{I}$ and
$(\mathrm{K}/2\pi) da^I$.
These are 3-form density-like and 2-form vorticity-like charge operators, respectively.
The gauge transformations are generated by these charge operators and are given by 
\begin{align}
a^{I}\to a^{I}+d\varphi^{I},
\quad
b^{I}\to b^{I}+d\zeta^{I}.
\label{quadratic gauge trsf}
\end{align}
Similar to the cubic theory discussed earlier,
demanding the invariance under \eqref{quadratic gauge trsf},
one can read off the conservation law of current, which is identical
to \eqref{current conservation}.
%
%Similar to the cubic theory discussed earlier, it seems that the coupling to currents is gauge non-invariant.
%The current coupling terms transform as
%\begin{align}
%&
%\delta_{IJ}a^{I}\wedge J_{qp}^{J} + \delta_{IJ}b^{I}\wedge J_{qv}^{J}
%\nonumber \\
%&
%\longrightarrow \delta_{IJ}a^{I}\wedge J_{qp}^{J}+ \delta_{IJ}b^{I}\wedge J_{qv}^{J}
%\nonumber \\
%&
% +d\varphi^{1}\wedge
% \left[J_{qp}^1+\frac{\mathrm{p}_2}{\mathrm{K}^2}d^{-1}J_{qv}^2\wedge J_{qv}^{1}
% -\frac{\mathrm{p}_1}{\mathrm{K}^2}d^{-1}J_{qv}^2\wedge J_{qv}^{2}\right]
%\nonumber \\
%& +d\varphi^{2}\wedge \left[J_{qp}^2
%+\frac{\mathrm{p}_1}{\mathrm{K}^2}d^{-1}J_{qv}^1\wedge J_{qv}^{2}
%-\frac{\mathrm{p}_2}{\mathrm{K}^2}d^{-1}J_{qv}^1\wedge J_{qv}^{1}\right]
%\nonumber \\
%& +\delta_{IJ}d\zeta^{I}\wedge J_{qv}^{J}
%\end{align}
%Hence by imposing gauge invariance, we recover the same conservation laws as before
%\begin{align}
%d\left[J_{qp}^1+\frac{p_2}{K^2}d^{-1}J_{qv}^2\wedge J_{qv}^{1}-\frac{p_1}{K^2}d^{-1}J_{qv}^2\wedge J_{qv}^{2}\right]=&\; 0 \nonumber \\
%d\left[J_{qp}^2+\frac{p_1}{K^2}d^{-1}J_{qv}^1\wedge J_{qv}^{2}-\frac{p_2}{K^1}d^{-1}J_{qv}^1\wedge J_{qv}^{1}\right]=&\; 0 \nonumber \\
%dJ_{qv}^{I}=&\; 0
%\end{align}
%\

\subsection{Three-loop braiding statistics}
\label{Three-loop braiding statistics}

To see the three-loop braiding statistics,
we need to quantize the coupled $BF$ theory (either the cubic theory or its quadratic avatar).
In this section, we consider the coupled $BF$ theory on topologically trivial spacetimes, e.g.,
$\Sigma=\mathbb{R}^3$, $\mathcal{M}=\mathbb{R}^3\times \mathbb{R}$,
and study the properties of the Wilson loop and Wilson surface operators.
In the next section, we put the coupled $BF$ theory on the spatial manifold with non-trivial topology,
the three torus, $\Sigma=T^3$.

As one of the simplest and quickest way to see the three-loop braiding statistics,
let us start by integrating over $a^I$ and $b^I$,
on both cubic and quadratic theories.
One then obtains
the effective action of the currents
\begin{align}
 \mathcal Z[J^I_{qp},J^J_{qv}]=e^{i S_{eff}[J_{qp}^I,J_{qv}^J]}=\int\mathcal{D}[a^I, b^I] e^{i S}
 \end{align}
 where
\begin{align}
S_{eff} &=
-\frac{2\pi}{\mathrm{K}}
\int (d^{-1}J^I_{qv})\wedge J^I_{qp}
\nonumber \\
\quad
&+
\frac{2\pi\mathrm{p}_1}{\mathrm{K}^3}
\int (d^{-1}J^1_{qv})\wedge (d^{-1}J^2_{qv})\wedge J^2_{qv}
\nonumber \\
\quad
&+
\frac{2\pi\mathrm{p}_{2}}{\mathrm{K}^3}
\int (d^{-1}J^2_{qv})\wedge (d^{-1}J^1_{qv})\wedge J^1_{qv}.
\end{align}
The first term in the effective action
describes, as in the ordinary $BF$ theory,
the quasparticle-quasivortex braiding statistics.
It is given in terms of the linking number of
\begin{align}
\mathrm{Link}( J_{qv}^I, J_{qp}^J)
=
\int_{\mathcal{M}}
(d^{-1}J^I_{qv})\wedge J^J_{qp},
\end{align}
in the spacetime $\mathcal{M}$.
On the other hand,
the second and third terms
include topological linking among three quasivortex loops,
i.e., three-loop braiding statistics.

%
%The equations of motion are
%\begin{align}
%&\frac{\mathrm{K}}{2\pi}da^{I}=J^{I}_{qv},
%\nonumber \\
%&
%\frac{\mathrm{K}}{2\pi}db^1+
%\frac{\mathrm{p}_1}{4\pi^2}a^2\wedge da^2
%\nonumber \\
%&\qquad
%-
%\frac{\mathrm{p}_2}{2\pi^2}
%a^2\wedge da^1
%+
%\frac{\mathrm{p}_2}{4\pi^2} da^2\wedge a^1= J_{qp}^1,
%\nonumber \\
%&
%\frac{\mathrm{K}}{2\pi}db^2+
%\frac{\mathrm{p}_2}{4\pi^2} a^1\wedge da^1
%\nonumber \\
%&\qquad
%-
%\frac{\mathrm{p}_1}{2\pi^2}a^1\wedge da^2
%+
%\frac{\mathrm{p}_1}{4\pi^2}da^1\wedge a^2= J_{qp}^2.
%\label{3loop EOM}
%\end{align}
%

%\subsubsection{The Wilson operator algebra}

The three-loop braiding statistics can also be discussed
by quantizing the theory and using the Wilson loop and Wilson surface operators.
Let us now take the quadratic theory \eqref{quadratic 3loopaction}.
From the coupling to the currents,
we read off the Wilson loop and Wilson surface operators
in the theory:
\begin{align}
A^{I}_L:=
\exp \left[i \int_L a^I\right],
\quad
W^{I}_S:=
\exp
\left[
i
\int_S
\Lambda^I
\right],
\label{surface op. quadratic theory}
\end{align}
where
$L$ and $S$ are arbitrary closed loop and surfaces in the spatial manifold $\Sigma$, respectively,
and
\begin{align}
\Lambda^I:=
b^I
+
\frac{\mathrm{q}_{\bar{I}}}{2\pi} a^I\wedge a^{\bar{I}}.
\end{align}
The commutation relations between these Wilson operators can be computed from the canonical commutation relation
\begin{align}
[a^I_{i}(x),b^{Jj}(y)]=\frac{{2\pi i}}{\mathrm{K}}
\delta^{IJ}
\delta_{i}^{j}\delta^{(3)}(x-y)
\end{align}
where
$a^I= a^I_i dx^i$,
$b^I= (1/2)b^I_{ij} dx^i \wedge dx^j$,
and
$
b^{I i}:=(1/2)\epsilon^{ijk}b^I_{jk}
$.
(We have adopted
the temporal gauge $a^I_0 = b^I_{0i}=0$.)
The exponents of the Wilson operators satisfy
\begin{align}
&
 \Big[ {\textstyle \int_C a^I},
 {\textstyle \int_S \Lambda^J} \Big]
 =
 \frac{2\pi i}{\mathrm{K}}
 \delta^{IJ} I(C,S),
 \nonumber \\
 &
 \Big[ {\textstyle \int_S \Lambda^I}, {\textstyle \int_{S'} \Lambda^J} \Big]
=
 \frac{2i}{\mathrm{K}^2}
\textstyle{\int_{S\#S'} }
 \left(
  \mathrm{p}_{\bar{J}} \delta^{I\bar{J}} a^{J}
 -
 \mathrm{p}_{\bar{J}} \delta^{IJ} a^{\bar{J}}
% \nonumber\\ &\quad \qquad
 \right)
\end{align}
and as before the repeated capital Roman indices are not summed over.
Here,
\begin{align}
I(C,S)= \int_{\Sigma} \delta(C)\wedge \delta(S)
\end{align}
is the intersection number between $C$ and $S$,
and $S\sharp S'$ is the intersection of $S$ and $S'$.

The three-loop braiding statistic is encoded in
the following product of Wilson operators
\cite{yoshida2015gapped}
\begin{align}
&
	(\hat{W}^{J\dag}_{S'}
	\hat{W}^{I\dag}_S
	\hat{W}^J_{S'}
	\hat{W}^I_S)
	\hat{W}^{K\dag}_{S^{\prime\prime}}
	(\hat{W}^{I\dag}_{S}
	\hat{W}^{J\dag}_{S'}
	\hat{W}^{I}_{S}
	\hat{W}^{J}_{S'})
	\hat{W}^{K}_{S^{\prime\prime}}
\nonumber \\
&\quad
=
\exp\left( \big[
\big[ i {\textstyle \int_S \hat{\Lambda}^I},
i {\textstyle \int_{S'}\hat{\Lambda}^J}\big],
i {\textstyle \int_{S^{\prime\prime}}
\hat{\Lambda}^{K}}
\big]
\right)
\end{align}
where
the triple commutator is given by
\begin{align}
&
 \Big[
 \Big[ {\textstyle \int_S \Lambda^I}, {\textstyle \int_{S'} \Lambda^J} \Big],
{\textstyle \int_{S^{\prime\prime}}\Lambda^K}
 \Big]
 \nonumber \\
 &\quad
 =
 \frac{4\pi \mathrm{p}_{\bar{J}} }{\mathrm{K}^3}
\big(
 \delta^{IJ}\delta^{\bar{J}K}
- \delta^{I\bar{J}}\delta^{JK}
 \big)
 I(S\#S', S^{\prime\prime}).
\end{align}
Physically, this product of Wilson operators braids loop $I$ with loop $J$ while both $I$ and $J$ are linked with `background' loop $K$.
Notice that the triple commutator satisfies the Jacobi identiy:
\begin{align}
&
\big[\big[
\textstyle{\int_{S}}\hat{\Lambda}^I,
\textstyle{\int_{S'}}\hat{\Lambda}^J\big],
\textstyle{\int_{S''}}\hat{\Lambda}^K \big]
+\big[\big[
\textstyle{\int_{S'}}\hat{\Lambda}^J,
\textstyle{\int_{S''}}\hat{\Lambda}^K\big],
\textstyle{\int_{S}}\hat{\Lambda}^I \big]
\nonumber \\
&
\quad
+\big[\big[
\textstyle{\int_{S''}}\hat{\Lambda}^K,
\textstyle{\int_{S}}\hat{\Lambda}^I\big],
\textstyle{\int_{S'}}\hat{\Lambda}^J \big]=0.
\end{align}
This is equivalent to the cyclic relation for the three-loop braiding phase first derived by Wang and Levin in
Ref.\ \onlinecite{wang2014braiding}.

\subsection{Quantization on a closed spatial manifold}
\label{Quantization on a closed spatial manifold}

In Sec.\ \ref{Three-loop braiding statistics},
the coupled $BF$ theory on topologically trivial spacetime is studied in the presence of
background quasiparticle and quasivortec currents.
In this section,
we consider spacetime wherein its spatial part $\Sigma$ is topologically non-trivial.
(Our setting closely parallels with Ref.\ \onlinecite{BergeronSemenoffSzabo1995}.)
In particular, we will focus on $\Sigma$ which is formal. (See the definition of
manifolds being formal below.) The simplest case is $\Sigma=T^3$.

\subsubsection{Mode decomposition and the zero-mode algebra}

We Hodge decompose the gauge fields as
$a^I$ and $b^I$ as
\begin{align}
a^I & =d\theta^I+\star dK^{I\prime}+\alpha^I_{l}\omega_{l},
\nonumber \\
b^I & =dK^I+\star d\theta^{I\prime}+\beta^I_{l}\eta_{l},
\end{align}
where
$d\theta^I, dK^I$
and
$\star dK^{\prime I}, \star d\theta^{I\prime}$
are the exact and coexact parts of the decomposition, respectively,
and
$\{\omega_l\}_l$ and $\{\eta_l\}_l$ are bases of
harmonic one- and two-forms, respectively.
$(d\omega_l = d\star \omega_l =0)$.
The ``zero modes'',
$\alpha^I_l$ and $\beta^I_l$,
which appear in the Hodge decomposition,
play a crucial role later.
Let
$\{ L^m\}$
and
$\{S^m\}$
be
a set of generators of
the first and second homology groups,
$H_1(\Sigma; \mathbb{Z})$
and
$H_2(\Sigma; \mathbb{Z})$,
respectively.
We define the linking matrix
by
\begin{align}
M^{mn}=
I(S^m, L^n)
%= \sum_{I_{mn}} \mathrm{sgn}\, (I_{mn}),
\end{align}
which counts
the signed intersections $I_{mn}$
of $S^m$ and $L^n$.
Furthermore,
\begin{align}
\int_{L^{m}}\omega_{l}=\delta_{l}^{m},
\quad
\int_{S^{m}}\eta_{l}=\delta_{l}^{m},
\quad
\int_{\Sigma} \omega_{l}\wedge\eta_{k}=M_{lk}
\end{align}
where $M_{lm}$ is the inverse of the linking matrix of $\Sigma$.

For the reason which will become clear momentarily,
we will work on a spatial manifold which is {\it formal}.
Here,
a Riemannian metric is called (metrically) formal if all
wedge products of harmonic forms are harmonic.
A closed manifold is called geometrically formal if it admits a formal
Riemannian metric.
\cite{kotschick2001products}
In particular, we will focus on the one of the simplest formal
manifolds; three-torus, $\Sigma=T^3$.

The Wilson loop/surface operators
for $L^m$ and $S^m$ on $\Sigma$
are written in terms of the zero modes,
$\alpha^I_l$ and $\beta^I_l$.
By noting
$
	\int_{L^i} a^I = M^{mi} \int_{\Sigma} a\wedge \eta_m = \alpha^I_i
$,
the Wilson operators for the gauge field $a^I_l$
are given by
\begin{align}
A^I_i := \exp i \int_{L^i} a^I = \exp i \alpha^I_i.
\end{align}
Similarly, one notes
$
	\int_{S^l} b^I = M^{lm} \int_{\Sigma} b\wedge \omega_m = \beta^I_l
$.
Since $\Sigma$ is formal,
\begin{align}
	\int_{S^l} a^I \wedge a^J
	&=
	\alpha^I_i \alpha^J_j \int_{S^l} \omega_i \wedge \omega_j
%	\nonumber \\
%	&=
%	\alpha^I_i \alpha^J_j
%	\int_{S^l}
%	C_{ijk} \eta_k
%	\nonumber \\
%	&=
	C_{ijl}
\end{align}
where the product of the two harmonic one-form
$\omega_i \wedge \omega_j $ is given in terms of the harmonic two-form as
$\omega_i \wedge \omega_j = C_{ijk}\eta_k$.
Thus, we consider the Wilson surface operators
\begin{align}
	W^I_i
	&:=
	\exp i \int_{S^i}
	\left(
	b^I + \frac{\mathrm{q}_{\bar{I}}}{2\pi}
	a^I\wedge a^{\bar{I}}
	\right)
	\nonumber \\
	&=
	\exp i
	\left(
	\beta^I_i + \frac{\mathrm{q}_{\bar{I}}}{2\pi}
C_{lmi}
\alpha^I_l \alpha^{\bar{I}}_m
	\right).
\end{align}

In the following,
we canonically quantize
the theory, and study the algebra obeyed by
the Wilson operators.
We will focus on $\Sigma=T^3$,
for which
the linking matrix
is simply the $3\times 3$
identity matrix,
\begin{align}
M^{mn}=\delta_{mn}.
\end{align}
We also take
\begin{align}
	C_{ijk} = \epsilon_{ijk}
\end{align}

Upon canonical quantization,
the zero modes,
$\hat{\alpha}^I_i$
and
$\hat{\beta}^I_i$,
now denoted with hat to indicate
they are quantum operators,
satisfy the commutator
\begin{align}
	\big[
	\hat{\alpha}^I_i, \hat{\beta}^J_j
	\big]
	=
	\frac{2\pi i}{\mathrm{K} } \delta_{ij} \delta^{IJ}.
\end{align}
Correspondingly,
we consider the set of Wilson operators
\begin{align}
&\quad
	\hat{A}^I_i = \exp\big( i \hat{\alpha}^I_i \big),
	\quad
	\hat{W}^I_i = \exp \big(i \hat{\Lambda}^I_i \big),
	\nonumber \\
	&
\mbox{where}
\quad
	\hat{\Lambda}^I_i =
	\hat{\beta}^I_i + \frac{\mathrm{q}_{\bar{I}}}{2\pi} \epsilon_{ijk}
	\hat{\alpha}^I_j \hat{\alpha}^{\bar{I}}_k,
%	\nonumber \\
%	\hat{\Lambda}^2_i &=
%	\hat{\beta}^2_i + \frac{\mathrm{q}_1}{2\pi} \epsilon_{ijk} \hat{\alpha}^2_j \hat{\alpha}^1_k
\end{align}
The commutators among $\hat{\alpha}^I_i$ and $\hat{\Lambda}^I_i$
are:
\begin{align}
	\big[
	\hat{\alpha}^I_i, \hat{\alpha}^J_j
	\big]
	&=
	0,
	\quad
	\big[
	\hat{\alpha}^I_i, \hat{\Lambda}^J_j
	\big]
	=
	\frac{2\pi i}{\mathrm{K} } \delta_{ij} \delta^{IJ}
	\nonumber \\
	\big[
	\hat{\Lambda}^I_i,
	\hat{\Lambda}^I_j
	\big]
	&=
	\frac{2 i \mathrm{q}_{\bar{I}}}{\mathrm{K}}\epsilon_{ijk} \hat{\alpha}^{\bar{I}}_k,
\nonumber \\
\big[
	\hat{\Lambda}^1_i ,
	\hat{\Lambda}^2_j
	\big]
	&=
	\frac{-i \mathrm{q}_2}{\mathrm{K}}\epsilon_{ijk} \hat{\alpha}^1_k
	-
	\frac{i \mathrm{q}_1}{\mathrm{K}}\epsilon_{ijk} \hat{\alpha}^2_k.
%	\nonumber \\
%\big[
%	\hat{\Lambda}^2_i ,
%	\hat{\Lambda}^2_j
%	\big]
%	&=
%	\frac{2 i \mathrm{q}_1}{\mathrm{K}}\epsilon_{ijk} \hat{\alpha}^1_k.
\label{commutators}
\end{align}
%It may be interesting to note that these commutators
%are somewhat similar to the Poisson brackets of
%the position and momentum operators
%of charged particles in the monopole background:
%\begin{align}
%	\{ x_i, x_j\} = 0,
%	\quad
%	\{x_i, p_j\}= \delta_{ij},
%	\quad
%	\{ p_i, p_j\}
%	= eg \epsilon_{ijk} \frac{x_k}{r^3},
%\end{align}
%where $e$ is the charge of the particles and $g$ the strength of the monopole.
%The corresponding symplectic two-form is given by
%\begin{align}
%	\Omega = dp_i \wedge dx_i
%	+
%	\frac{eg}{2} \epsilon_{ijk} \frac{x_k}{r^3} dx_i \wedge dx_j
%\end{align}

\subsubsection{The Wilson operator algebra and three-loop braiding statistics}
The three-loop braiding phase can be read off from the algebra of Wilson surface operators. To compute the algebra of Wilson operators, we use the
Baker-Campbell-Hausdorff formula:
\begin{align}
	e^{\hat{A}} e^{\hat{B}}
	&=
	\exp\Big(
	\hat{A}+\hat{B}+\frac{1}{2}[\hat{A},\hat{B}]
	\nonumber \\
	&\qquad
	+
	\frac{1}{12}\left[ \hat{A}-\hat{B}, \left[\hat{A},\hat{B}\right] \right]+\cdots
	\Big).
\end{align}
Thus, for the products of Wilson operator,
\begin{align}
&
	\hat{W}^{I\dag}_i \hat{W}^{J\dag}_j \hat{W}^I_i \hat{W}^J_j
\nonumber \\
	&\quad =
	\exp\left( [i\hat{\Lambda}^I_i, i\hat{\Lambda}^J_j ]\right),
	\nonumber \\
	&
	(\hat{W}^{J\dag}_j
	\hat{W}^{I\dag}_i
	\hat{W}^J_j
	\hat{W}^I_i)
	\hat{W}^{K\dag}_k
	(\hat{W}^{I\dag}_i
	\hat{W}^{J\dag}_j
	\hat{W}^{I}_i
	\hat{W}^{J}_j)
	\hat{W}^{K}_k
\nonumber \\
&\quad =
\exp\left( [
[ i \hat{\Lambda}^I_i, i\hat{\Lambda}^J_j ],
i\hat{\Lambda}^{K}_k
]
\right).
\label{products of Wilson operators}
\end{align}
The triple commutator is a phase and the above algebra of Wilson surface operator describes the three-loop braiding phase.
This is consistent with previous work on three-loop braiding statistics.
\cite{yoshida2015gapped}
To have a non-zero three-loop braiding phase,
$I,J,K$ cannot be all equal.
$i,j,k$ cannot be all equal neither.
We list non-zero triple-linking phase factors below:
\begin{align}
	\big[\big[\hat{\Lambda}^I_i,\hat{\Lambda}^I_j \big], \hat{\Lambda}^{\bar{I}}_k \big]
	&=
	-\epsilon_{ijk} \frac{4\pi \mathrm{q}_{\bar{I}}}{\mathrm{K}^2},
	\nonumber \\
	\big[\big[\hat{\Lambda}^1_i,\hat{\Lambda}^2_j \big], \hat{\Lambda}^1_k \big]
	&=
	\big[\big[\hat{\Lambda}^2_i,\hat{\Lambda}^1_j \big], \hat{\Lambda}^1_k \big]
	=
	\epsilon_{ijk} \frac{2\pi \mathrm{q}_2}{\mathrm{K}^2},
	\nonumber \\
	\big[\big[\hat{\Lambda}^1_i,\hat{\Lambda}^2_j \big], \hat{\Lambda}^2_k \big]
	&=
	\big[\big[\hat{\Lambda}^2_i,\hat{\Lambda}^1_j \big], \hat{\Lambda}^2_k \big]
	=
	\epsilon_{ijk} \frac{2\pi \mathrm{q}_1}{\mathrm{K}^2}.
%	\nonumber \\
%	\big[\big[\hat{\Lambda}^2_i,\hat{\Lambda}^2_j \big], \hat{\Lambda}^1_k \big]
%&=
%	-\epsilon_{ijk} \frac{4\pi \mathrm{q}_1}{\mathrm{K}^2}.
\end{align}
%Notice that the triple commutator satisfies the Jacobi identiy:
%\begin{align}
%\big[\big[\hat{\Lambda}^I_i,\hat{\Lambda}^J_ j\big], \hat{\Lambda}^K_k \big]+\big[\big[\hat{\Lambda}^J_j,\hat{\Lambda}^K_ k\big], \hat{\Lambda}^I_i \big]+\big[\big[\hat{\Lambda}^K_k,\hat{\Lambda}^I_ i\big], \hat{\Lambda}^J_j \big]=0
%\end{align}
%This is equivalent to the cyclic relation for the three-loop braiding phase first derived by Wang and Levin in \cite{wang2014braiding}
%

\subsubsection{Large gauge invariance}
\label{Large gauge invariance two flavors}

Unlike the infinitesimal gauge transformations, the large gauge transformations cannot be derived from the conserved charges or Gauss law constraints of the action. 
However,
the large gauge invariance can be deduced by demanding 
the invariance of the Wilson operators $\hat{A}^I_i$ and $\hat{W}^I_i$. 
(Or vice versa: once the large gauge transformations are properly defined, 
the Wilson operators are defined as those that are invariant under the large 
gauge transformations.)
Hence the correct large gauge transformations are
\begin{align}
\hat{\alpha}^I_i &\to \hat{\alpha}'^{I}_i=\hat{\alpha}^I_i + 2\pi n^I_i,
\nonumber \\
\hat{\beta}^I_i
&\to \hat{\beta}'^{I}_i=
\hat{\beta}^I_i -
\mathrm{q}_{\bar{I}} \epsilon_{ijk} \left[
n^I_j \hat{\alpha}^{\bar{I}}_k
+
\hat{\alpha}^I_j n^{\bar{I}}_k
+
2\pi
n^I_j n^{\bar{I}}_k
\right].
%\nonumber \\
%\hat{\beta}^2_i
%&\to \hat{\beta}'^2_i=\hat{\beta}^2_i -
%\mathrm{q}_1 \epsilon_{ijk} \left[
%n^2_j \hat{\alpha}^1_k
%+
%\hat{\alpha}^2_j n^1_k
%+
%2\pi n^2_j n^1_k
%\right].
\label{large_gauge}
\end{align}
%After large gauge transformation,
%\begin{align}
%\hat{\beta}^{\prime 1}_i
%&= \hat{\beta}^1_i -
%\mathrm{q}_2 \epsilon_{ijk} \left[ n^1_j \hat{\alpha}^2_k + \hat{\alpha}^1_j n^2_k + 2\pi n^1_j n^2_k \right],
%\nonumber \\
%\hat{\beta}^{\prime 2}_i
%&= \hat{\beta}^2_i -
%\mathrm{q}_1 \epsilon_{ijk} \left[ n^2_j \hat{\alpha}^1_k + \hat{\alpha}^2_j n^1_k + 2\pi n^2_j n^1_k \right],
%\end{align}
It is worth noticing that, 
since $\hat{\beta}_i^I$ transforms non-linearly under large gauge transformations,
the $\big[\hat{\beta}^{I}_i,\hat{\beta}^{J}_j\big]$ commutator is not preserved.
In fact,
\begin{align}
	\big[
	\hat{\beta}^{\prime  I}_i,
	\hat{\beta}^{\prime I}_j
	\big]
	%&=
%\left[
%	\hat{\beta}^{1}_i,
%\mathrm{q}_2 \epsilon_{jpq} \hat{\alpha}^1_p n^2_q
%	\right]
	%+
%\left[
%\mathrm{q}_2 \epsilon_{irs} \hat{\alpha}^1_r n^2_s ,
%	\hat{\beta}^{1}_j
%	\right]
%	\nonumber \\
%&	=
%\mathrm{q}_2 \epsilon_{jpq}  n^2_q
%\left[
%	\hat{\beta}^{1}_i,
 %\hat{\alpha}^1_p
%	\right]
%	+
%\mathrm{q}_2 \epsilon_{irs} n^2_s
%\left[
%\hat{\alpha}^1_r,
%	\hat{\beta}^{1}_j
%	\right]
%	\nonumber \\
%	&=
%\mathrm{q}_2 \epsilon_{jpq}  n^2_q
%\frac{2\pi i}{\mathrm{K}} \delta_{ip}
%	+
%\mathrm{q}_2 \epsilon_{irs} n^2_s
%\frac{-2\pi i}{\mathrm{K}} \delta_{rj}
%\nonumber \\
%&=
%\mathrm{q}_2 \epsilon_{jiq}  n^2_q
%\frac{2\pi i}{\mathrm{K}}
%	+
%\mathrm{q}_2 \epsilon_{ijs} n^2_s
%\frac{-2\pi i}{\mathrm{K}}
%\nonumber \\
%&=
%\frac{2\pi i \mathrm{q}_2}{\mathrm{K}}
%\left(
 %\epsilon_{jiq}  n^2_q
%	-
 %\epsilon_{ijs} n^2_s
 %\right)
& = -\frac{4\pi i \mathrm{q}_{\bar{I}}}{\mathrm{K}}\epsilon_{ijk}n_{k}^{\bar{I}},
\nonumber \\
%\frac{ i 2\mathrm{q}_2}{\mathrm{K}}
%\epsilon_{ijk}
%(2\pi  n^2_k)
\big[\hat{\beta}'^{1}_i,\hat{\beta}'^{2}_j\big]&=
\frac{2\pi i}{\mathrm{K}}\epsilon^{ijk}\left(\mathrm{q}_1n_k^2+\mathrm{q}_2n_k^1\right),
%\nonumber \\
%\big[\hat{\beta}'^{2}_i,\hat{\beta}'^{2}_j\big]&= -\frac{4\pi i \mathrm{q}_1}{\mathrm{K}}\epsilon_{ijk}n_{k}^1.
\end{align}
However the algebra of observables, i.e the Wilson algebra transforms covariantly under large gauge transformations.
E.g.,
\begin{align}
	\big[
	\hat{\Lambda}^{\prime 1}_i,
	\hat{\Lambda}^{\prime 1}_j
	\big]
%&	=
%	\frac{-2 i \mathrm{q}_2}{\mathrm{K}}\epsilon_{ijk} \hat{\alpha}^2_k
%-
%\frac{ i 2\mathrm{q}_2}{\mathrm{K}}
%\epsilon_{ijk}
%(2\pi  n^2_k)
%\nonumber \\
&= \frac{-2 i \mathrm{q}_2}{\mathrm{K}}\epsilon_{ijk} (\hat{\alpha}^2_k + 2\pi n^2_k)
= \frac{-2 i \mathrm{q}_2}{\mathrm{K}}\epsilon_{ijk} \hat{\alpha}^{\prime 2}_k.
%\nonumber \\
%%\left[
%	\hat{\Lambda}^{\prime 2}_i,
%	\hat{\Lambda}^{\prime 2}_j
%	\right]
%&= \frac{-2 i \mathrm{q}_1}{\mathrm{K}}\epsilon_{ijk} \hat{\alpha}^{\prime 1}_k \nonumber \\
%\left[
%	\hat{\Lambda}^{\prime 2}_i,
%	\hat{\Lambda}^{\prime 2}_j
%	\right]
%&=\frac{i\epsilon_{ijk}}{K}\left(q_1\hat{\alpha}_k^{\prime 2}+ q_2\hat{\alpha}_k^{\prime 1}\right)
\end{align}
Therefore,
the operator algebra is preserved under the large gauge transformations.

As for the Wilson operators 
$\hat{A}^I_i$ and $\hat{W}^I_i$, 
they are invariant under the large gauge transformations
\eqref{large_gauge} by construction.
Nevertheless, it should be noted that their product may not be so, 
as seen 
in 
$\hat{W}^{I\dag}_i\hat{W}^{J\dag}_j\hat{W}^I_i\hat{W}^J_j$
in Eq.\ \eqref{products of Wilson operators}
(note the commutators in Eq.\ \eqref{commutators}),
although the algebra of the Wilson operators is gauge covariant;
The algebra of the Wilson operators generated by 
$\hat{A}^I_i$ and $\hat{W}^I_i$
and
that generated by
$\hat{A}^{I\prime}_i$ and $\hat{W}^{I\prime}_i$ 
are isomorphic.
While 
$\hat{W}^{I\dag}_i\hat{W}^{J\dag}_j\hat{W}^I_i\hat{W}^J_j$
is not gauge invariant, 
the product
$
(\hat{W}^{J\dag}_j
	\hat{W}^{I\dag}_i
	\hat{W}^J_j
	\hat{W}^I_i)
	\hat{W}^{K\dag}_k
	(\hat{W}^{I\dag}_i
	\hat{W}^{J\dag}_j
	\hat{W}^{I}_i
	\hat{W}^{J}_j)
	\hat{W}^{K}_k
	$
and the triple commutator 
	$
[
[ i \hat{\Lambda}^I_i, i\hat{\Lambda}^J_j ],
i\hat{\Lambda}^{K}_k
]
$
are large gauge invariant, and so is the three-loop braiding phase. 

\subsection{Wave function in terms of Wilson operators}
\label{Wave function in terms of Wilson operators}

In the previous section, we have computed the algebra of the Wilson operators
of the coupled $BF$ theory
for non-contractible loops and surfaces on $T^3$.
As we will show momentarily, in this section,
we can build and label
%the ground state Hilbert space in terms of the Wilson operators, and label
all the ground states on $T^3$ in terms of these Wilson operators.
Furthermore,
we will use these ground states to calculate the modular $\mathcal{T}$ and $\mathcal{S}$ matrices,
which encode the spin and the braiding statistics of topological excitations.
\cite{WangLevin2015, WangWen2015, JiangMesarosRan2014, WangLevin2014,wang2016quantum, wan2015twisted}

For this purpose, it is advantageous
to construct the three-dimensional version of minimum entropy states (MESs),
which are a special choice of the basis for the ground state multiplet.\cite{Zhang2012}
By calculating the overlap between MESs before and after applying the modular $\mathcal{S}$ and $\mathcal{T}$ transformations,
we can read off the braiding statistics for particle-loop and three-loop braiding.
The MES basis has been constructed before
in Refs.\ \onlinecite{WangLevin2015, WangWen2015, JiangMesarosRan2014, WangLevin2014} for
microscopic models defined on lattices.
We will show that the $\mathcal{S}$ and $\mathcal{T}$ matrices that we are going to calculate
are the same as that for their model,
and therefore we verify that our model is the continuum version of the Dijkgraaf-Witten model.
\cite{dijkgraaf1990topological}
These $\mathcal{S}$ and $\mathcal{T}$ matrices are also consistent with those calculated from
the partition functions of the gapless boundary theory in our previous paper.
\cite{chen2015bulk}

\subsubsection{The ordinary $BF$ theory}

Before we study
the $\mathcal{S}$ and $\mathcal{T}$ matrices for the coupled $BF$ theory,
as a warm up,
we first demonstrate our strategy for the ordinary $BF$ theory on $T^3$.
The zero modes of the $BF$ theory obey the commutation relation
$
[\hat{\alpha}_i,\hat{\beta}_j]=\delta_{ij}({2\pi i}/{\mathrm{K}})
$,
$[\hat{\alpha}_i,\hat{\alpha}_j]=[\hat{\beta}_i,\hat{\beta}_j]=0$,
where $i,j=1,2,3$.
The Wilson loop and surface operators for non-contractible loops and surfaces on $T^3$ are given by
$\hat{A}_i=\exp(i\hat{\alpha}_i)$ and $\hat{B}_i=\exp(i\hat{\beta}_i)$, and by taking powers thereof.
They satisfy
\begin{align}
\hat{A}_i\hat{B}_j=\delta_{ij}e^{-2\pi i/\mathrm{K}}\hat{B}_j\hat{A}_i.
\end{align}

We define and choose a vacuum state (a reference state) $|0\rangle$ such that all $\hat{A}_i$'s are diagonal.
All the other ground states can be generated, starting from $|0\rangle$, by applying $\hat{B}_i$:
$\hat{B}_3^a\hat{B}_2^b\hat{B}_1^c|0\rangle$.
These states are the eigenstate of $\hat{A}_i$ operator.
The $\mathcal{S}$ and $\mathcal{T}$ matrices for this basis is the Kronecker delta and
do not tell us the information about the spin and braiding statistics at all.
To extract the spin and braiding statistics,
we construct the three-dimensional version of MESs in $z$-direction
by considering
the eigenstates of the Wilson operators $\hat{A}_1$, $\hat{A}_2$ and $\hat{B}_3$.
Namely, we consider the set of states given by
\begin{align}
|\Psi_{n_1,n_2,n_3}\rangle=\frac{1}{\sqrt{\mathrm{K}}}\sum_{\lambda}e^{\frac{2\pi i\lambda n_3}{\mathrm{K}}}
\hat{B}_3^{\lambda}\hat{B}_1^{n_1}\hat{B}_2^{n_2}|0\rangle
\end{align}
where $\lambda,n_1,n_2,n_3\in \mathbb{Z}_{\mathrm{K}}$.
As we check momentarily,
the $\mathcal{T}$ matrix acts diagonally on these states
-- an expected feature for states with definite ``topological'' or ``anyonic'' charge.

The $\mathcal{T}$ transformation can be visualized as the shear deformation in the $xz$ plane (as its two-dimensional counter part on $T^2$).
Hence, under the $\mathcal{T}$ transformation, $\hat{B}_1\to\hat{B}_1\hat{B}_3$.
The MESs $|\Psi_{n_i}\rangle$ are transformed under $\mathcal{T}$ as
\begin{align}
\mathcal{T}|\Psi_{n_i}\rangle&=\frac{1}{\sqrt{\mathrm{K}}}\sum_{\lambda}e^{\frac{2\pi i\lambda n_3}{\mathrm{K}}}\hat{B}_3^{\lambda+n_1}\hat{B}_1^{n_1}\hat{B}_2^{n_2}|0\rangle\nonumber\\
&=e^{-\frac{2\pi i n_1n_3}{\mathrm{K}}}|\Psi_{n_i}\rangle.
\end{align}
Therefore, $\mathcal{T}$ matrix takes a diagonal form for the MESs,
and encodes information related to a (3+1)d analogue of topological spin.

The modular $\mathcal{S}$ transformation is slightly more non-trivial and can be decomposed into $\mathcal{S}_{13}$ and $\mathcal{S}_{12}$,
which are $90^{\circ}$ rotation in the $xz$ and $xy$ planes, respectively.
Under the $\mathcal{S}_{13}$ transformation,
\begin{align}
\mathcal{S}_{13}|\Psi_{n_i}\rangle
=\frac{1}{\sqrt{\mathrm{K}}}\sum_{\lambda}
e^{\frac{2\pi i\lambda n_3}{\mathrm{K}}}\hat{B}_1^{-\lambda}\hat{B}_3^{n_1}\hat{B}_2^{n_2}|0\rangle.
\end{align}
Therefore, the $\mathcal{S}_{13}$ matrix for the MES basis is calculated as
\begin{align}
\langle\Psi_{n_i^{\prime}}|\mathcal{S}_{13}|\Psi_{n_i}\rangle
&=\frac{1}{\mathrm{K}}\sum_{\lambda^{\prime},\lambda}\langle 0|e^{\frac{2\pi i}{\mathrm{K}}(-\lambda^{\prime}n_3^{\prime}+\lambda n_3)}
\nonumber \\
&\quad
\times
\hat{B}_2^{-n_2^{\prime}}\hat{B}_3^{-n_1^{\prime}}
\hat{B}_1^{\lambda^{\prime}}\hat{B}_3^{\lambda}\hat{B}_1^{n_1}\hat{B}_2^{n_2} |0\rangle\nonumber\\
&=\frac{1}{\mathrm{K}}\delta_{n^{\ }_2,n_2^{\prime}}e^{\frac{2\pi i}{\mathrm{K}}(n^{\ }_1n_3^{\prime}+n_1^{\prime}n_3)}.
\end{align}
In the above derivation, we use $-\lambda^{\prime}=n_1$, $\lambda=n_1^{\prime}$ and $n_2=n_2^{\prime}$.
Combined with the $\mathcal{S}_{12}$ transformation, we can write down the modular $\mathcal{S}$ matrix
\begin{align}
\mathcal{S}
=\frac{1}{\mathrm{K}}\delta_{n_1,n_2^{\prime}}e^{-\frac{2\pi i}{\mathrm{K}}(n_3^{\prime}n_2-n_3n_1^{\prime})}.
\end{align}

We can easily generalize the above results to the two decouple copies of $BF$ theories on $T^3$.
The commutators among zero modes are
\begin{align}
[\hat{\alpha}_i^I,\hat{\beta}_j^J]=\delta_{ij}\delta_{IJ}\frac{2\pi i}{\mathrm{K}},
\quad [\hat{\alpha}_i^I,\hat{\alpha}_j^J]=[\hat{\beta}_i^I,\hat{\beta}_j^J]=0.
\label{two_bf_comm}
\end{align}
The MES basis is given by
\begin{align}
|\Psi_{n_i}^{l_i}\rangle&=\frac{1}{\mathrm{K}}\sum_{\lambda_1,\lambda_2}
e^{\frac{2\pi i}{\mathrm{K}}(\lambda_1 n_3+\lambda_2 l_3)}
(\hat{B}_3^1)^{\lambda_1}(\hat{B}_1^1)^{n_1}(\hat{B}_2^1)^{n_2}
\nonumber\\
&\quad \times (\hat{B}_3^2)^{\lambda_2}(\hat{B}_1^2)^{l_1}(\hat{B}_2^2)^{l_2}
|0\rangle.
\end{align}
These states are an eigenstate of
$\hat{A}_1^I$, $\hat{A}_2^I$, $\hat{B}_3^I$
($I=1,2$).
The modular $\mathcal{T}$ and $\mathcal{S}$ matrices are given by
\begin{align}
\mathcal{T}&=
\delta_{n^{\ }_i,n_i^{\prime}}\delta_{l_i,l_i^{\prime}}e^{-\frac{2\pi i }{\mathrm{K}}(n_1n_3+l_1l_3)}
\nonumber \\
\mathcal{S}
&=\frac{1}{\mathrm{K}^2}\delta_{n^{\ }_1,n_2^{\prime}}
\delta_{l^{\ }_1,l_2^{\prime}}
e^{-\frac{2\pi i}{\mathrm{K}}(n_3^{\prime}n_2-n_3n_1^{\prime})-\frac{2\pi i}{\mathrm{K}}(l_3^{\prime}l_2-l_3l_1^{\prime})}
\end{align}

\subsubsection{The coupled $BF$ theory:
wave functions in terms of $\hat{\alpha}$ and $\hat{\beta}$}

For the coupled $BF$ theory realizing three-loop braiding statistics defined in Eq.\ \eqref{quadratic 3loopaction},
the commutators between $\alpha_i^I$ and $\beta_i^I$
are identical to those in the two decoupled copies of $BF$ theories defined in Eq.\ \eqref{two_bf_comm}.
On the other hand, if we consider $\Lambda_i^I$ instead of $\beta_i^I$, the commutators are
\begin{align}
[\hat{\alpha}_i^I,\hat{\Lambda}_j^J]=\delta_{ij}\delta_{IJ}\frac{2\pi i}{\mathrm{K}},\quad [\hat{\alpha}_i^I,\hat{\alpha}_j^J]=0,\quad [\hat{\Lambda}_i^I,\hat{\Lambda}_j^J]\neq 0
\end{align}
In the next two subsections, we will construct two sets of MESs in terms of $\beta_i^I$ and $\Lambda_i^I$.

Let us first construct MESs using $\hat{\beta}^i_I$.
Similar to the two decoupled copies of $BF$ theories,
$\hat{A}_1^I$, $\hat{A}_2^I$, $\hat{B}_3^I$ ($I=1,2$)
commute with each other. Therefore, we define the eigenstate for these operators as
\begin{align}
&|\Psi_{n_i}^{l_i}\rangle\sim \sum_{\lambda_1,\lambda_2}
e^{\frac{2\pi i\lambda_1}{\mathrm{K}}
\left(n_3+\frac{l\times n}{\mathrm{K}}\right)
+\frac{2\pi i\lambda_2}{\mathrm{K}}\left(l_3+\frac{n\times l}{K}\right)}
\nonumber\\
&\times (\hat{B}_1^1)^{n_1}(\hat{B}_2^1)^{n_2} (\hat{B}_1^2)^{l_1}(\hat{B}_2^2)^{l_2}  (\hat{B}_3^1)^{\lambda_1} (\hat{B}_3^2)^{\lambda_2} |0\rangle,
\label{gauge inv. states}
\end{align}
where $l\times n :=l_1n_2-l_2n_1$.
One verifies that the states constructed in Eq.\  \eqref{gauge inv. states}
are invariant under the large gauge transformations \eqref{large_gauge},
up to a phase factor (which can depend on $n_i$ and $l_i$).
Here for simplicity, we consider $\mathrm{q}_1=\mathrm{q}_2=1$.
Different from the two decoupled copies of $BF$ theories,
we require $\lambda_1,\lambda_2, n_i,l_i\in \mathbb{Z}_{\mathrm{K}^2}$ and $n_3$ ($l_3$)
is shifted by $(l\times n)/\mathrm {K}$ ($(n\times l)/\mathrm{K}$).
Because of the extra factor of $\mathrm{K}^{-1}$ in $l\times n/\mathrm{K}$ or $n\times l/\mathrm{K}$,
it may seem that there are $\mathrm{K}^{12}$ different eigenstates, as opposed to $\mathrm{K}^6$, which is
the expected number of ground states for two copies of $BF$ theories.
This is however not the case once we properly
reorganize these wave functions.
%, we will show there are only $\mathrm{K}^6$ independent states.
Let us introduce
\begin{align}
&n_1\equiv \mathrm{K} t_1+\bar{n}_1,\quad
n_2\equiv \mathrm{K}t_2+\bar{n}_2\nonumber\\
&\bar{n}_3\equiv n_3+s\times\bar{n}+\bar{l}\times t+\bar{l}_2(s_1-t_1)\ \mbox{mod}\ \mathrm{K}
\nonumber\\
&l_1\equiv \mathrm{K}s_1+\bar{l}_1,\quad
l_2\equiv \mathrm{K}s_2+\bar{l}_2\nonumber\\
&\bar{l}_3\equiv l_3+t\times\bar{l}+\bar{n}\times s+\bar{n}_2(t_1-s_1)\ \mbox{mod}\ \mathrm{K}
\end{align}
where $\bar{n}_i,\bar{l}_i, t_1,t_2,s_1,s_2\in \mathbb{Z}_{\mathrm{K}}$.
In terms of these quantum numbers,
the wave functions depend only on (and are labeled by)
$\bar{n}_i$ and $\bar{l}_i$,
as they can be written as
\begin{align}
&|\Psi_{\bar{n}_i}^{\bar{l}_i}\rangle=
\frac{1}{\mathrm{K}^3}
\sum_{\lambda_{1,2},t_{1,2}, s_{1,2}}
e^{\frac{2\pi i\lambda_1}{\mathrm{K}}
\left(\bar{n}_3+\frac{\bar{l}\times \bar{n}}{\mathrm{K}}\right)
+\frac{2\pi i\lambda_2}{\mathrm{K}}\left(\bar{l}_3+\frac{\bar{n}\times \bar{l}}{\mathrm{K}}\right)}
\nonumber\\
&\quad
\times  (\hat{B}_1^1)^{\mathrm{K}t_1+\bar{n}_1}(\hat{B}_2^1)^{\mathrm{K}t_2+\bar{n}_2}\nonumber\\
&\quad
\times (\hat{B}_1^2)^{\mathrm{K}s_1+\bar{l}_1}(\hat{B}_2^2)^{\mathrm{K}s_2+\bar{l}_2}
(\hat{B}_3^1)^{\lambda_1} (\hat{B}_3^2)^{\lambda_2} |0\rangle.
\label{wave1}
\end{align}
This construction of the ground states is analogous to the construction of the surface partition functions realizing
the three-loop braiding phase in Ref.\ \onlinecite{chen2015bulk}.

With respect to the ground states \eqref{wave1}, 
the $\mathcal{T}$ transformation is diagonal, 
\begin{align}
\mathcal{T}|\Psi_{\bar{n}_i}^{\bar{l}_i}\rangle
=
e^{-\frac{2\pi i}{\mathrm{K}}\bar{n}_1\left( \bar{n}_3+\frac{\bar{l} \times \bar{n}}{\mathrm{K}} \right)
-\frac{2\pi i}{\mathrm{K}}\bar{n}_1\left( \bar{n}_3+\frac{\bar{l} \times \bar{n}}{\mathrm{K}} \right) }
 |\Psi_{\bar{n}_i}^{\bar{l}_i}\rangle
\label{three_loop_t}
\end{align}
On the other hand,
the $\mathcal{S}_{13}$ matrix is
\begin{align}
&
\langle \Psi_{\bar{n}_i^{\prime}}^{\bar{l}_i^{\prime}} |
\mathcal{S}_{13}
|\Psi_{\bar{n}_i}^{\bar{l}_i}\rangle
=\frac{1}{\mathrm{K}^6}
\sum_{
\lambda^{\ }_{1,2},
\lambda_{1,2}^{\prime},
t^{\ }_{1,2},
t_{1,2}^{\prime},
s^{\ }_{1,2},s_{1,2}^{\prime}}
\nonumber\\
&\quad
\times
\langle 0|
e^{
-\frac{2\pi i\lambda_1^{\prime}}{\mathrm{K}}\left(\bar{n}_3^{\prime}+\frac{\bar{l}^{\prime}\times \bar{n}^{\prime}}{\mathrm{K}}\right)
-\frac{2\pi i\lambda_2^{\prime}}{\mathrm{K}}\left(\bar{l}_3^{\prime}+\frac{\bar{n}^{\prime}\times \bar{l}^{\prime}}{\mathrm{K}}\right)
}
\nonumber \\
&\quad
\times
e^{
+\frac{2\pi i\lambda_1}{\mathrm{K}} \left(\bar{n}_3+\frac{\bar{l}\times \bar{n}}{\mathrm{K}}\right)
+\frac{2\pi i\lambda_2}{\mathrm{K}}\left(\bar{l}_3+\frac{\bar{n}\times \bar{l}}{\mathrm{K}}\right)
}
\nonumber\\
&\quad
\times (\hat{B}_2^2)^{-(\mathrm{K}s_2^{\prime}+\bar{l}_2^{\prime})}
(\hat{B}_3^2)^{-(\mathrm{K}s_1^{\prime}+\bar{l}_1^{\prime})}(\hat{B}_1^2)^{\lambda_2^{\prime}}
\nonumber\\
&\quad
\times (\hat{B}_2^1)^{-(\mathrm{K}t_2^{\prime}+\bar{n}_2^{\prime})}
(\hat{B}_3^1)^{-(\mathrm{K}t_1^{\prime}+\bar{n}_1^{\prime})}(\hat{B}_1^1)^{\lambda_1^{\prime}}
\nonumber\\
&\quad
\times (\hat{B}_3^1)^{\lambda_1}(\hat{B}_1^1)^{\mathrm{K}t_1+\bar{n}_1}(\hat{B}_2^1)^{\mathrm{K}t_2+\bar{n}_2}
\nonumber\\
&\quad
\times (\hat{B}_3^2)^{\lambda_2}(\hat{B}_1^2)^{\mathrm{K}s_1+\bar{l}_1}(\hat{B}_2^2)^{\mathrm{K}s_2+\bar{l}_2} |0\rangle
\nonumber\\
&
=\frac{1}{\mathrm{K}^2}\delta_{\bar{n}_2,\bar{n}_2^{\prime}}\delta_{\bar{l}_2,\bar{l}_2^{\prime}}e^{i\theta^{\bar{n}_i^{\prime},\bar{l}_i^{\prime},
\bar{n}_i,\bar{l}_i}}
\end{align}
where $\theta^{\bar{n}_i^{\prime},\bar{l}_i^{\prime},
\bar{n}_i,\bar{l}_i}$ is given by
\begin{align}
&\theta^{\bar{n}_i^{\prime},\bar{l}_i^{\prime},\bar{n}_i,\bar{l}_i}=
\frac{2\pi}{\mathrm{K}}(\bar{n}_3\bar{n}_1^{\prime}+\bar{n}_3^{\prime}\bar{n}_1+\bar{l}_3\bar{l}_1^{\prime}+\bar{l}_3^{\prime}\bar{l}_1)
\nonumber\\
&\quad
+\frac{2\pi}{\mathrm{K}^2}\left[ (\bar{l}\times\bar{n})(n_1^{\prime}-l_1^{\prime})+(\bar{l}^{\prime}\times\bar{n}^{\prime})(n_1-l_1) \right].
\label{three_loop_phase}
\end{align}
From the calculation of $\mathcal{S}_{13}$, we can further calculate the modular $\mathcal{S}$ matrix,
\begin{align}
\mathcal{S}
&=\frac{1}{\mathrm{K}^2}
\delta_{n_1,n_2^{\prime}}\delta_{l_1,l_2^{\prime}}
e^{-\frac{2\pi i}{\mathrm{K}}(n_3^{\prime}n_2-n_3n_1^{\prime})-\frac{2\pi i}{\mathrm{K}}(l_3^{\prime}l_2-l_3l_1^{\prime})}
\nonumber\\
&\quad
\times e^{-\frac{2\pi i}{\mathrm{K}^2}\left[ (n_1+l_1)(n_2l_1^{\prime}+n_1^{\prime}l_2)-2n_2n_1^{\prime}l_1-2n_1l_2l_1^{\prime} \right]}.
\label{three_loop_s}
\end{align}
The $\mathcal{S}$ and $\mathcal{T}$ matrices obtained in this way are the same as those obtained for
the surface partition functions in our previous work,
and other bulk calculations.
\cite{chen2015bulk, WangLevin2015, WangWen2015, JiangMesarosRan2014, WangLevin2014}

\subsubsection{The coupled $BF$ theory: wave function in terms of $\hat{\alpha}$ and $\hat{\Lambda}$}

While we have succeeded, by using $B^I_i$,
in constructing the ground state wave functions and in computing
the $\mathcal{T}$ and $\mathcal{S}$ matrices,
it is also worth trying to use $\Lambda^I_i$ instead of $B^I_i$ to construct wave functions.
One motivation for this is that $\exp i \Lambda^I_i$ are the Wilson surface operators, while $\exp i \beta^I_i$ are not.
Although $\Lambda_{i}^I$ may not commute with each other,
$\hat{A}_1^1$, $\hat{A}_2^1$, $\hat{W}_3^1$, $\hat{A}_1^2$, $\hat{A}_2^2$ and $\hat{W}_3^2$ still commute with each other,
and we can write down the eigenstates for them,
\begin{align}
&|\Psi_{n_i}^{l_i}\rangle=
\frac{1}{\mathrm{K}}\sum_{\lambda_{1,2}}
e^{\frac{2\pi i}{\mathrm{K}} (\lambda_1n_3 + \lambda_2l_3) }
\nonumber\\
&\quad \times (\hat{W}_3^1)^{\lambda_1}(\hat{W}_3^2)^{\lambda_2}(\hat{W}_1^1)^{n_1}(\hat{W}_1^2)^{l_1}(\hat{W}_2^1)^{n_2}(\hat{W}_2^2)^{l_2}|0\rangle
\label{wave2}
\end{align}
where $\lambda_1,\lambda_2,n_i,l_i\in\mathbb{Z}_{\mathrm{K}}$.
Since $\hat{W}^I_i$ do not mutually commute,
the ordering of $\hat{W}_i^I$ is important
when generating a set of wave functions.
We choose this particular order so that $\mathcal{S}$ and $\mathcal{T}$
matrices are the same as those calculated in the previous subsection.
Notice that since $\hat{W}_i^I$ is invariant under the large gauge transformations,
so is this wave function.

The matrix elements of $\mathcal{S}_{13}$ can be calculated as
\begin{align}
&\langle\Psi_{n_i^{\prime}}^{l_i^{\prime}}|\mathcal{S}_{13}|\Psi_{n_i}^{l_i}\rangle=
\frac{1}{\mathrm{K}^2}\sum_{\lambda_{1,2}, \lambda_{1,2}^{\prime}}
e^{
\frac{2\pi i}{\mathrm{K}}
( -\lambda_1^{\prime}n_3^{\prime} -\lambda_2^{\prime}l_3^{\prime} +\lambda_1n_3 +\lambda_2l_3 )
}
\nonumber\\
&\quad
\times
\langle 0|
(\hat{W}_2^2)^{-l_2^{\prime}}(\hat{W}_2^1)^{-n_2^{\prime}}(\hat{W}_3^2)^{-l_1^{\prime}}(\hat{W}_3^1)^{-n_1^{\prime}}(\hat{W}_1^2)^{\lambda_2^{\prime}}(\hat{W}_1^1)^{\lambda_1^{\prime}}\nonumber\\
&\quad
\times (\hat{W}_3^1)^{\lambda_1}(\hat{W}_3^2)^{\lambda_2}(\hat{W}_1^1)^{n_1}(\hat{W}_1^2)^{l_1}(\hat{W}_2^1)^{n_2}(\hat{W}_2^2)^{l_2}|0\rangle\nonumber\\
&
=\frac{1}{\mathrm{K}^2}\delta_{n_2,n_2^{\prime}}\delta_{l_2,l_2^{\prime}}
e^{i\theta^{n_i^{\prime},n_i,l_i^{\prime},l_i}},
\end{align}
where $\theta^{n_i^{\prime},n_i,l_i^{\prime},l_i}$ is the same as that in Eq.\ \eqref{three_loop_phase}.
One can then check that the modular $\mathcal{S}$ matrix also matches with the previous calculation
in terms of $B^I_i$, Eq.\ \eqref{three_loop_s}.

As for the $\mathcal{T}$ transformation,
since $\hat{W}^I_1$ and $\hat{W}^I_3$ do not commute with each other,
their transformation properties under the $\mathcal{T}$ transformation are more complicated.
%we need to calculate the transformation of Wilson operators under $\mathcal{T}$ transformation more carefully.
Using the knowledge that $\Lambda=\beta+\alpha\times\alpha$,
we decompose $\hat{W}_i^I$ as
\begin{align}
\hat{W}_i^I=\hat{B}_i^I\hat{C}_i^I
\quad
\mbox{where}\quad
\hat{C}_i^I=
\exp\left(\frac{i\mathrm{q}_{\bar{I}}}{2\pi}\epsilon_{ijk}\hat{\alpha}_j^I\hat{\alpha}_k^{\bar{I}}\right).
\end{align}
We propose that under the $\mathcal{T}$ transformation,
\begin{align}
(\hat{B}_1^I)^{n_1}(\hat{C}_1^I)^{n_1}
\to
(\hat{B}_1^I)^{n_1}(\hat{B}_3^I)^{n_1}(\hat{C}_1^I)^{n_1}(\hat{C}_3^I)^{n_1}.
\end{align}
The above result can be rewritten in terms of the $\hat{W}_i^I$ operators as
\begin{align}
(\hat{W}_1^I)^{n_1}\to (\hat{W}_3^I)^{n_1/2}(\hat{W}_1^I)^{n_1}(\hat{W}_3^I)^{n_1/2}.
\end{align}
According to this definition,
under the $\mathcal{T}$ transformation, $|\Psi_{n_i}^{l_i}\rangle$ are transformed as
%Let us now consider the $\mathcal{T}$ transformation.
%In the ordinary $BF$ theory, under the $\mathcal{T}$ transformation, $(\hat{B}_1)^{n_1}\to (\hat{B}_1)^{n_1}(\hat{B}_3)^{n_1}$.
%However, since $\hat{W}_1$ and $\hat{W}_3$ do not commute with each other,
%there is ambiguity in the definition of $\mathcal{T}$ transformation,
%we postulate that under $\mathcal{T}$ transformation,
%$(\hat{W}_1)^{n_1}\to (\hat{W}_3)^{n_1/2}(\hat{W}_1)^{n_1}(\hat{W}_3)^{n_1/2}$.
%If so,
%under the $\mathcal{T}$ transformation, $|\Psi_{n_i}^{l_i}\rangle$ are transformed as
\begin{align}
&\mathcal{T}|\Psi_{n_i}^{l_i}\rangle
=\frac{1}{\mathrm{K}}\sum_{\lambda_{1,2}}
e^{
\frac{2\pi i}{\mathrm{K}} (\lambda_1n_3 + \lambda_2l_3)
}
\nonumber\\
&\quad
\times (\hat{W}_3^1)^{\lambda_1}(\hat{W}_3^2)^{\lambda_2}(\hat{W}_3^1)^{n_1/2}(\hat{W}_1^1)^{n_1}(\hat{W}_3^1)^{n_1/2}\nonumber\\
&\quad
\times (\hat{W}_3^2)^{l_1/2} (\hat{W}_1^2)^{l_1}(\hat{W}_3^2)^{l_1/2}(\hat{W}_2^1)^{n_2}(\hat{W}_2^2)^{l_2}|0\rangle.
\end{align}
Therefore,
the $\mathcal{T}$ matrix is given by
\begin{align}
%\langle\Psi_{n_i^{\prime}}^{l_i^{\prime}}|\mathcal{T}|\Psi_{n_i}^{l_i}\rangle&=\delta_{n_i,n_i^{\prime}}\delta_{l_i,l_i^{\prime}}
\mathcal{T}&=\delta_{n_i,n_i^{\prime}}\delta_{l_i,l_i^{\prime}}
e^{ -\frac{2\pi i}{\mathrm{K}}(n_3n_1+l_3l_1)-\frac{2\pi i}{\mathrm{K}^2}(l\times n)(n_1-l_1)}.
\end{align}
This also matches with the previous calculation Eq.\ \eqref{three_loop_t}.

%In the above two subsections,
%we consider two basis of ground state wave functions Eqs.\ \eqref{wave1} and \eqref{wave2}.
%If we use the knowledge that $\hat{\Lambda}\equiv\hat{\beta}+\hat{\alpha}\times \hat{\alpha}$,
%and plug it into Eq.\ \eqref{wave2}, we expect that these two wave functions are describing the same state.
%

\section{Four-loop braiding theory}
\label{Four-loop braiding theory}

\subsection{The quartic theory}

In this section, we consider the following $BF$ theory with quartic coupling:
\begin{align}
S&=\int_{\mathcal{M}}
\Bigg[\frac{\mathrm{K}}{2\pi}\delta_{IJ}b^I\wedge da^J
+\epsilon_{IJKL}\frac{\mathrm{p}}{4!} a^I\wedge a^J\wedge a^K\wedge a^L
\nonumber \\
&\qquad
-\delta_{IJ}b^I\wedge J_{qv}^J-\delta_{IJ}a^I\wedge J_{qp}^J
\Bigg],
%\nonumber \\
%&=\int_{\mathcal{M}}
%\Bigg[\frac{\mathrm{K}}{2\pi}\delta_{IJ}b^I\wedge da^J
%+\mathrm{p} a^1\wedge a^2\wedge a^3\wedge a^4
%\nonumber \\
%&\quad
%-\delta_{IJ}b^I\wedge J_{qv}^J-\delta_{IJ}a^I\wedge J_{qp}^J
%\Bigg]
\label{action_quartic}
\end{align}
where $I,J\in 1,2,3,4$.
This action can be considered as describing a discrete (lattice) gauge theory
with the gauge group
$\mathbb{Z}_{\mathrm{K}}\times \mathbb{Z}_{\mathrm{K}}\times \mathbb{Z}_{\mathrm{K}}\times \mathbb{Z}_{\mathrm{K}}$.
$\mathrm{p}$ is a parameter of the theory, and is given by
\begin{align}
\mathrm{p}=\frac{\mathrm{q} \mathrm{K}^3}{(2\pi)^3},
\quad
\mathrm{q}=0,1,\ldots,\mathrm{K}-1.
\label{p_def}
\end{align}

We will show that the quartic theory \eqref{action_quartic} realizes 
non-trivial four-loop braiding statistics. 
Our reasoning presented below parallels our discussion on three-loop braiding statistics 
realized in the cubic theory. 

\subsubsection{Equations of motion}
The first term in the action \eqref{action_quartic}
describes the particle-loop braiding process, as in the ordinary $BF$ theory. 
On the other hand, as we will discuss,
the second term describes four-loop braiding process.
To develop understanding of the four-loop braiding process, 
let us first write down the equations of motion
\begin{align}
&\frac{\mathrm{K}}{2\pi}da^I=J_{qv}^I, \nonumber\\
&\frac{\mathrm{K}}{2\pi}db^I+\frac{\mathrm{p}}{6}  \epsilon_{IJKL} a^J\wedge a^K\wedge a^L=J_{qp}^I.
%&\frac{\mathrm{K}}{2\pi}db^1+\mathrm{p}  a^2\wedge a^3\wedge a^4=J_{qp}^1\nonumber\\
%&\frac{\mathrm{K}}{2\pi}db^2-p a^1\wedge a^3\wedge a^4=J_{qp}^2\nonumber\\
%&\frac{\mathrm{K}}{2\pi}db^3+p a^1\wedge a^2\wedge a^4=J_{qp}^3\nonumber\\
%&\frac{\mathrm{K}}{2\pi}db^4+p a^1\wedge a^2\wedge a^3=J_{qp}^4
\label{eom_bulk}
\end{align}
Let us consider a fixed static quasiparticle and quasivortex configuration and integrate the equation of motion over space.
By solving the first equation of motion as $a^I=(2\pi/\mathrm{K})(d^{-1}J_{qv}^I)$,
plugging the solution to the other equations of motion, and integrating over space $\Sigma$,
\begin{align}
&
\frac{\mathrm{K}}{2\pi}\int_{\Sigma} db^I=
\int_{\Sigma}
J_{qp}^I
\nonumber \\
&\quad
-
\frac{\mathrm{q}}{6}
\epsilon_{IJKL}
\int_{\Sigma}(d^{-1}J_{qv}^J)\wedge (d^{-1}J_{qv}^K)\wedge (d^{-1}J_{qv}^L).
%\frac{\mathrm{K}}{2\pi}\int_{\Sigma} db^1&=-\mathrm{q}\int_{\Sigma}(d^{-1}J_{qv}^2)\wedge (d^{-1}J_{qv}^3)\wedge (d^{-1}J_{qv}^4)+J_{qp}^1\nonumber\\
%\frac{\mathrm{K}}{2\pi}\int_{\Sigma} db^2&=+ \mathrm{q} \int_{\Sigma}(d^{-1}J_{qv}^1)\wedge (d^{-1}J_{qv}^3)\wedge (d^{-1}J_{qv}^4)+J_{qp}^2\nonumber\\
%\frac{\mathrm{K}}{2\pi}\int_{\Sigma} db^3&=-\mathrm{q} \int_{\Sigma}(d^{-1}J_{qv}^1)\wedge (d^{-1}J_{qv}^2)\wedge (d^{-1}J_{qv}^4)+J_{qp}^3\nonumber\\
%\frac{\mathrm{K}}{2\pi}\int_{\Sigma} db^4&=+\mathrm{q}\int_{\Sigma}(d^{-1}J_{qv}^1)\wedge (d^{-1}J_{qv}^2)\wedge (d^{-1}J_{qv}^3)+J_{qp}^4
\label{eom_quartic}
\end{align}
%The second term on the right-hand side of the above equation comes from  
%\begin{align}
%&
%\int_{\Sigma} a^I\wedge a^J\wedge a^K
%\nonumber \\
%&\quad
%=\left(\frac{2\pi}{\mathrm{K}}\right)^3\int_{\Sigma} (d^{-1}J_{qv}^I)\wedge (d^{-1}J_{qv}^J)\wedge (d^{-1}J_{qv}^K).
%\end{align}
The second term on the right-hand side of the above equation  
comes from 
\begin{align}
&
\mathrm{Borr}(J^I_{qv},J^J_{qv},J^K_{qv})
\nonumber \\
&\quad = \int_{\Sigma} 
(d^{-1}J_{qv}^I)\wedge (d^{-1}J_{qv}^J)\wedge (d^{-1}J_{qv}^K).
\end{align}
and involves three quasivortex loops. If any two of them are mutually unlinked, i.e., $d(a^I\wedge a^J)=0$,
this term describes the triple linking number of the Borromean ring configuration and is a topological invariant.
\cite{Berger1990, 2014JPhCS.544a2014B}
As in the three-loop braiding theory, 
the equation of motion \eqref{eom_quartic},
suggests the Borromean ring `dresses' the $I$-th quasiparticle
(Fig.\ \ref{fig:fig0}).

To see that $\mathrm{Borr}$ is a topological invariant,
let us introduce $g^K=\epsilon^{IJK}a^I\wedge a^J$. 
If we require that any two of the flux loops are mutually unlinked, i.e., $d(a^I\wedge a^J)=0$,
this constraint leads to $g^K=du^K$, where $u^{K}$ is a one-form gauge field and describes the effective magnetic flux loop 
formed by $a^{I}$ and $a^J$.
Then, $\mathrm{Borr}$ can be written as  
\begin{align}
&
\int_{\Sigma} a^1\wedge a^2\wedge a^3
\nonumber \\
&
=\int_{\Sigma}  a^1\wedge du^1
=\int_{\Sigma}  a^2 \wedge du^2
=\int_{\Sigma}  a^3\wedge du^3.
\end{align}
This is equivalent to a Chern-Simons integral and describes the Hopf linking number between $da^k$ and $du^k$.

\subsubsection{Gauge invariance}

In the absence of sources,
there are two sets of gauge transformations that leave the action invariant:
The usual 1-form gauge transformation
\begin{align}
b^{I}\to b^{I}+d\zeta^{I},
\quad a^{I}\to a^{I}
\end{align}
and a shifted 0-form gauge transformation
\begin{align}
a^{I}\to a^{I}+d\varphi^{I}, \quad b^{I}\to b^{I}+\frac{\pi \mathrm{p}}{3\mathrm{K}}\epsilon^{IJKL}(a^{J}\wedge a^{K})\varphi^{L}.
\end{align}
Formally, these transformations can be read off by identifying the operators that generate the Gauss law constraints.

Similar to the three-loop braiding theory described earlier,
it seems that the coupling to currents is gauge non-invariant.
However, by demanding gauge invariance, we can read off the topological currents.
The terms with coupling to sources transform under the 0-form gauge transformations as
\begin{align}
&
a^{I}\wedge J_{qp}^{I}+b^{I}\wedge J_{qv}^{I}
\\
&\longrightarrow
a^{I}\wedge J_{qp}^{I}+b^{I}\wedge J_{qv}^{I}
\nonumber \\
&\quad
-\varphi^{I}d\left[J_{qp}^{I}
-\epsilon^{IJKL}\frac{\mathrm{q}}{6}d^{-1}J_{qv}^{J}\wedge d^{-1}J_{qv}^{K}\wedge d^{-1}J_{qv}^{L} \right].
\nonumber
\end{align}
Hence we can read off the current conservation law
\begin{align}
d\left[J_{qp}^{I}-\epsilon^{IJKL}
\frac{\mathrm{q}}{6}d^{-1}J_{qv}^{J}\wedge d^{-1}J_{qv}^{K}\wedge d^{-1}J_{qv}^{L} \right]=0.
\label{conservation law four-loop}
\end{align}
If all the pair of quasivortex loops are mutually unlinked, the first term on the right side describes the triple linking number for the borromean ring configuration.
The above equations then indicate,
as the equation of motion \eqref{eom_quartic}, 
that the effective particle comes from two parts, 
the real particle excitation and the Borromean ring configuration.
On the other hand, the 1-form gauge symmetry furnishes the second `ordinary' conservation law $dJ_{qv}^{I}=0$.

\subsubsection{Four-loop braiding statistics}

\begin{figure}[bt]
\centering
\includegraphics[scale=.4]{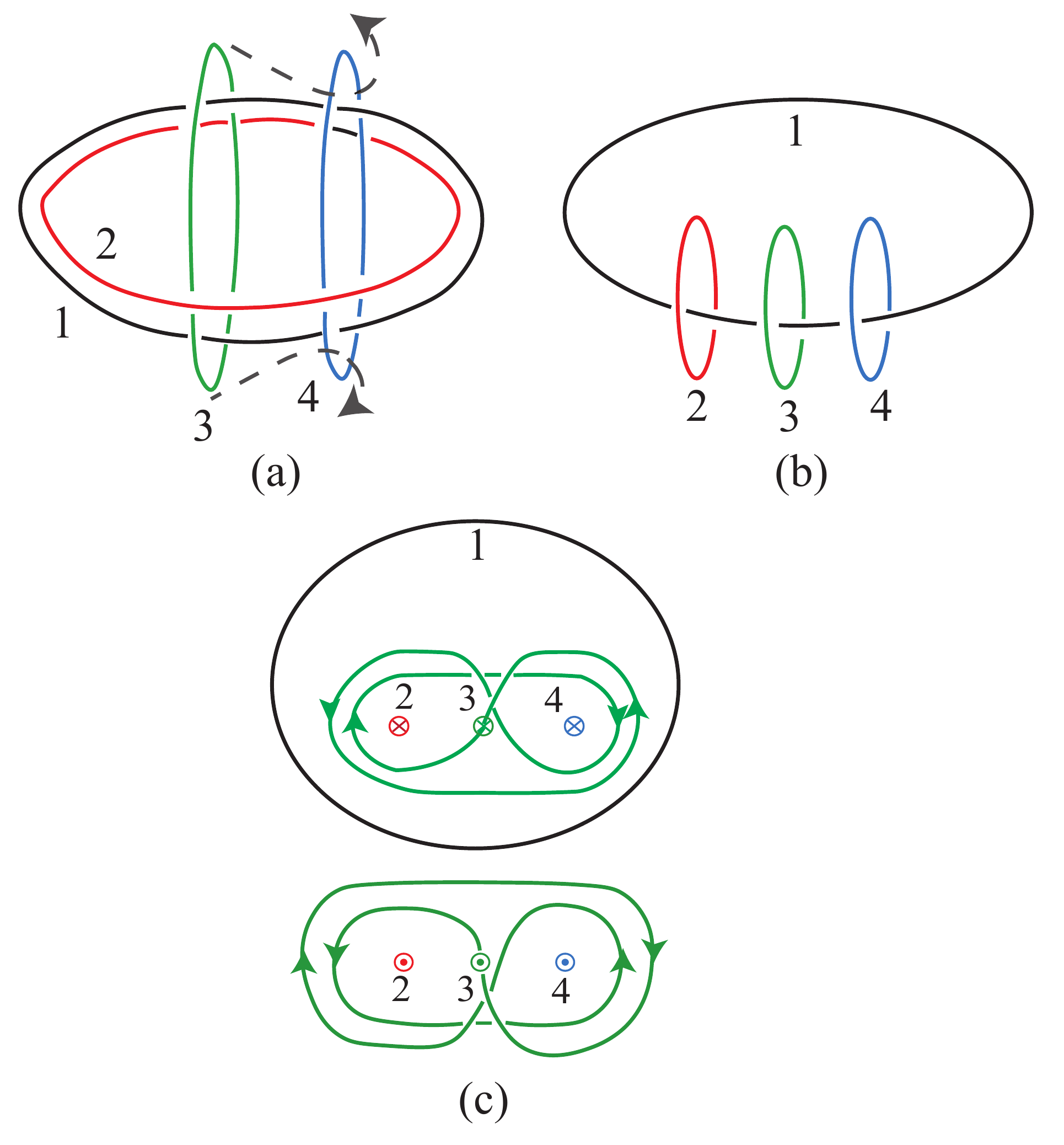}
\caption{
Four-loop braiding process in $3+1$ dimensions.
(a)
Loop 1, 2 and 3 form a Borromean ring configuration.
Similarly,  loop 1, 2 and 4 form a Borromean ring configuration.
Alternatively,
$L_1$ and $L_2$ form an effective loop $L_{12}$, with loop 3 and 4 are linked to $L_{34}$.
In this case, it is the same as the three loop braiding process with $L_{12}$ as the base loop.
Braiding $L_3$ around $L_4$ gives rise to a non-trivial phase $2\pi n_1n_2n_3n_4/\mathrm{K}$.
(b) Loop $L_1$ is the base loop. $L_2$, $L_3$ and $L_4$ are linked with $L_1$. 
This braiding process can be understood by dimensional reduction to the $(2+1)$ dimensions in Fig.\ \ref{fig:fig1} (c).
}
\label{fig:fig1}
\end{figure}

That the Borromean ring configuration can be treated as an effective particle,
as seen from the equation of motion \eqref{eom_quartic}
and the conservation law \eqref{conservation law four-loop}
suggests the theory may realize non-trivial statistics involving 
four loop-like excitations (four-loop braiding statistics).
%suggests 
%the braiding phase which is quartic in terms of the quantum number of loop excitations, $\sim n_1 n_2 n_3 n_4/\mathrm{K}$.
%Here,
%a loop excitation labeled by the quantum number $n_1\in \mathbb{Z}_{\mathrm{K}}$
%is braided with a Borromean ring which depends on three quantum numbers
%$n_{2,3,4}\in \mathbb{Z}_{\mathrm{K}}$.
%Alternatively, this four-loop braiding process can be understood through the analogue with three-loop braiding process.
Following three-loop braiding process, 
%Alternatively, this four-loop braiding process can be understood through the analogue with three-loop braiding process.
we postulate the four-loop braiding process as shown in Fig.\ \ref{fig:fig1}.
%As shown in Fig.\ \ref{fig:fig1}, 
In Fig.\ \ref{fig:fig1}, 
we consider the loop $L_1$ and $L_2$ form an effective base loop $L_{12}$,
with loop 3 and 4 are linked to $L_{12}$.
Braiding $L_3$ around $L_4$ gives rise to a non-trivial phase $\sim  n_1n_2n_3n_4/\mathrm{K}$.
Furthermore, we can also understand this braiding process by treating loop $L_1$ as an base loop, 
with loops $L_2$, $L_3$ and $L_4$ linked to $L_1$ (Fig.\ \ref{fig:fig1} (a)). 
Loop $L_2$ braids around $L_3$ and $L_4$. 
We will verify this argument shortly  by computing the algebra of Wilson operators. 

\begin{figure}[bt]
\centering
\includegraphics[scale=.35]{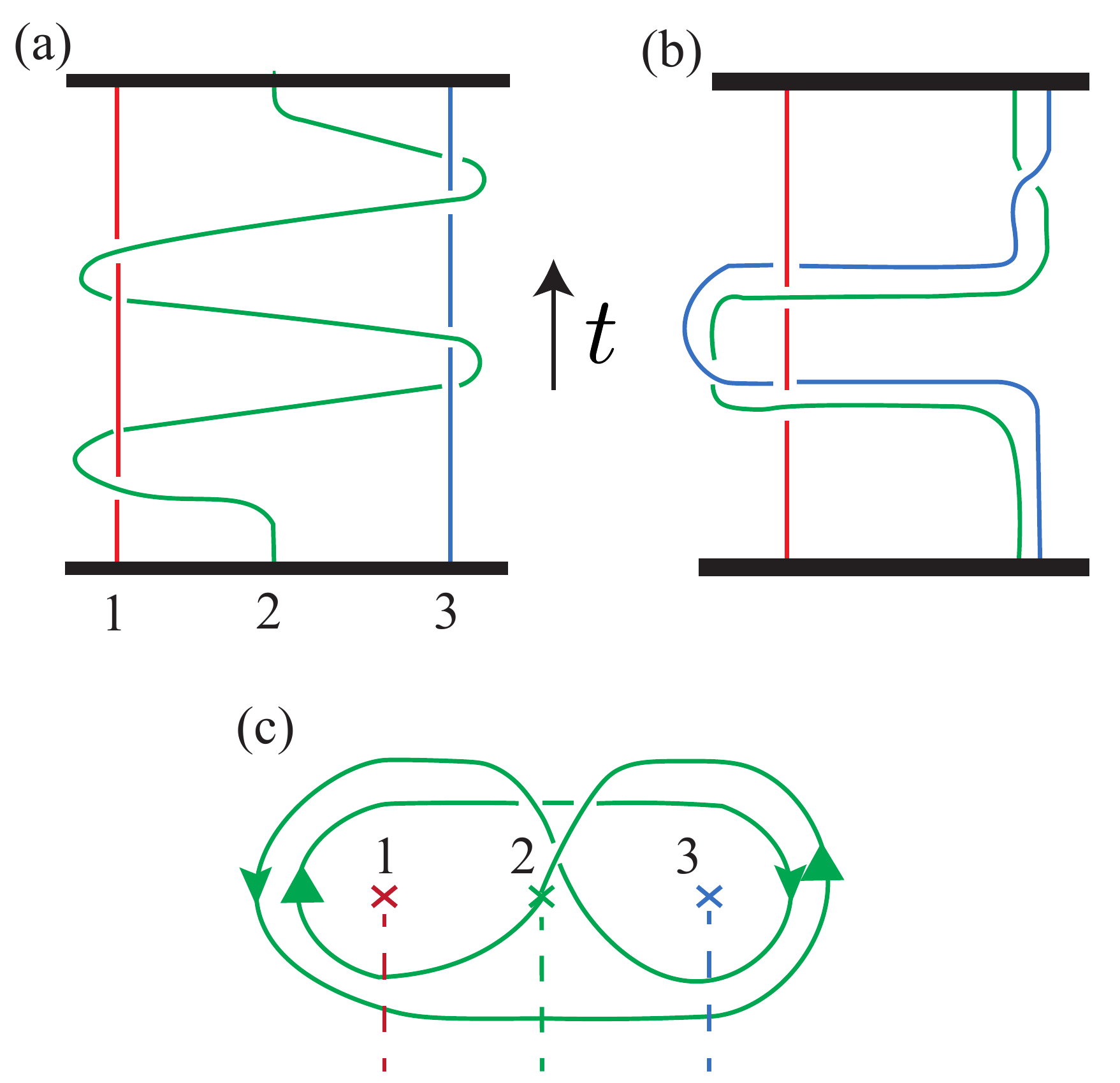}
\caption{
Three-particle braiding in $2+1$ dimensions. 
In (a), particle 1, 2 and 3 are labeled by three different colors red, green and blue. 
We braid particle 2 around 1 and 3 four times. The Wilson loops for particles 1, 2 and 3 are mutually unlinked. 
For instance, if there is no Wilson loop for particle 3, the braiding between 1 and 2 is trivial. 
Nevertheless, the three Wilson loops 1, 2 and 3 together form a Borromean ring in $2+1$ dimensions. 
In (b), we treat particle 2 and 3 as an effective particle and braid it around particle 1.
This process is topologically equivalent to (a).
(c) is the projection of (a) to the two dimensional spatial plane. 
The braiding of particle 2 around 1 is trivial if there is no particle 3.}
\label{fig:fig2}
\end{figure}

The last point of view can be better understood by 
considering dimensional reduction to one lower dimension as in Fig.\ \ref{fig:fig1} (c).
The dimensional reduction of the $(3+1)$ dimensional quartic theory
leads to the following $(2+1)$ dimensional cubic theory, 
\begin{align}
S=\int_{\mathcal{M}}\left[\frac{\mathrm{K}}{2\pi}\delta_{IJ}b^I\wedge da^J+\mathrm{p} a^1\wedge a^2\wedge a^3\right]
\label{action_cubic}
\end{align}
where $I,J=1,2,3$, $b^I$ and $a^I$ are one-form,
and $\mathrm{p}$ equals to $\mathrm{p}=\mathrm{q}\mathrm{K}^2/(2\pi)^2$
where $\mathrm{q}=0,1,\ldots,\mathrm{K}-1$.
The first term is the $BF$ theory and is related to the Hopf linking number for the particle current loops in 
$(2+1)$ dimensions,
which describes the particle-particle braiding process.
For the second term, if any two of particle current loops are mutually unlinked,
it is the Borromean ring and describes the braiding process involving three particles.
This braiding process has been discussed in Ref.\ \onlinecite{WangLevin2015} and can be understood as in Fig.\ \ref{fig:fig2}.

\subsection{The quadratic theory}

As we did for the coupled $BF$ theories realizing the three-loop braiding,
we can also consider an alternative quartic theory
instead of the quartic theory.
Let us consider:
\begin{align}
S &= \frac{\mathrm{K}}{2\pi} \int \delta_{IJ} b^I\wedge da^I
\nonumber \\
&\quad 
- \int \delta_{IJ} a^I \wedge J^J_{qp}
- \int \delta_{IJ}\Lambda^{I}\wedge J_{qv}^{J}
%\left[ b^I + \epsilon_{IJKL} \frac{\mathrm{p}}{4} d^{-1} (a^J\wedge a^K \wedge a^L) \right]
%\wedge J^M_{qv}
%\nonumber \\
%&\quad
%- \int \left[ b^1 - \frac{\mathrm{p}}{4} d^{-1} (a^2\wedge a^3 \wedge a^4) \right]\wedge J^1_{qv}
%\nonumber \\
%&\quad
%- \int \left[ b^2 + \frac{\mathrm{p}}{4} d^{-1} (a^1\wedge a^3 \wedge a^4) \right]\wedge J^2_{qv}
%\nonumber \\
%&\quad
%- \int \left[ b^3 - \frac{\mathrm{p}}{4} d^{-1} (a^1\wedge a^2 \wedge a^4) \right]\wedge J^3_{qv}
%\nonumber \\
%&\quad
%- \int \left[ b^4 + \frac{\mathrm{p}}{4} d^{-1} (a^1\wedge a^2 \wedge a^3) \right]\wedge J^4_{qv}
\end{align}
where $I,J=1,\ldots, 4$ and
\begin{align}
\Lambda^{I}&:=b^{I}-\frac{\mathrm{p}}{3!}\epsilon^{IJKL} d^{-1}\left(a^J\wedge a^K\wedge a^L\right).
%\nonumber \\
%B^{1}:=&\;b^{1}-\mathrm{p}d^{-1}\left(a^2\wedge a^3\wedge a^4\right) \nonumber \\
%B^{2}:=&\;b^{2}+\mathrm{p}d^{-1}\left(a^1\wedge a^3\wedge a^4\right) \nonumber \\
%B^{3}:=&\;b^{3}-\mathrm{p}d^{-1}\left(a^1\wedge a^2\wedge a^4\right) \nonumber \\
%B^{4}:=&\;b^{4}+\mathrm{p}d^{-1}\left(a^1\wedge a^2\wedge a^3\right)
\end{align}
The equations of motion are the same as Eq.\ \eqref{eom_quartic}.
Here,
the precise meaning of the term $\int d^{-1}(a^J\wedge a^K\wedge a^L)\wedge J^I_{qv}$
can be understood by taking $J_{qv}= \delta(S)$, which gives rise to for example
$
\int_{S} [b^1 -\mathrm{p} d^{-1}(a^2\wedge a^3\wedge a^4)]
$.
Looking for a volume $V$ which satisfies
$\partial V=S$,
this can be written as
$
\int_{S} b^1 - \mathrm{p} \int_V(a^2\wedge a^3\wedge a^4)
$.

Using the quadratic theory,
let us now discuss the algebra of the Wilson operators.
The canonical commutators are the same as the ordinary $BF$ theory
and hence
$
\left[{\textstyle \int_C} a^I, {\textstyle \int_S} \Lambda^J\right]=(2\pi i/\mathrm{K})\delta^{IJ} I(C, S).
$
On the other hand,
the multiple commutators among $\int_{S}\Lambda^I$ are
%\begin{align}
%&\;\left[ \int_{S_1} \Lambda^1, \int_{S_2} \Lambda^2\right]\nonumber\\
%=&\;\left[\int_{x\in\Sigma\times \mathbb R}\epsilon^{\mu\nu\lambda\rho}\left(\frac{1}{2}b_{\mu\nu}^{1}\frac{1}{2}\delta_{\lambda\rho}(S_1)-\mathrm{p}a_{\mu}^{2}a_{\nu}^{3}a_{\lambda}^{4}\delta_{\rho}(V_1)\right), \right. \nonumber \\
%&\;\left.  \int_{y\in\Sigma\times \mathbb R}\epsilon^{\alpha\beta\theta\phi}\left(\frac{1}{2}b_{\alpha\beta}^{2}\frac{1}{2}\delta_{\theta\phi}(S_2)+\mathrm{p}a_{\alpha}^{3}a_{\beta}^{4}a_{\theta}^{1}\delta_{\phi}(V_2)\right)
%\right] \nonumber \\
%=&\; \frac{2\pi i\mathrm{p}}{K}\int_{x\in \Sigma\times \mathbb R}\epsilon^{i\alpha\beta\lambda}\Big[\delta_{i0}(S_1)\delta_{\alpha}(V_2)-\delta_{i0}(S_2)\delta_{\alpha}(V_1)
%\Big]a_{\beta}^{3}a_{\lambda}^{4}
%\nonumber
%\end{align}
\begin{align}
&
\left[ \textstyle{\int_{S_1}} \Lambda^I, \textstyle{\int_{S_2}} \Lambda^J\right]
\nonumber \\
& \qquad
=
(-\mathrm{p}\epsilon^{IJPQ})
\left[\frac{-2\pi i}{\mathrm{K}}\right]
%\frac{12\pi i \mathrm{p}}{\mathrm{K}}
%\int_{S_1\sharp V_2 + S_2\sharp V_1} a^P\wedge a^Q,
\int_{\partial(V_1\sharp V_2)} a^P\wedge a^Q,
\nonumber \\
&
\left[\left[ \textstyle{\int_{S_1}} \Lambda^I, \textstyle{\int_{S_2}} \Lambda^J\right], \textstyle{\int}_{S_3}\Lambda^K\right]
\nonumber \\
& \qquad
=
(-4\mathrm{p}\epsilon^{IJKQ})
\left[\frac{-2\pi i}{\mathrm{K}}\right]^2
%\frac{96\pi^2 \mathrm{p}}{\mathrm{K}^2}
%\int_{(S_1\sharp V_2 + S_2\sharp V_1)\sharp S_3} a^Q,
\int_{\partial (V_1 \sharp V_2)\sharp S_3} a^Q,
\nonumber \\
&
\left[
\left[\left[ \textstyle{\int_{S_1}} \Lambda^I, \textstyle{\int_{S_2}} \Lambda^J\right], \textstyle{\int}_{S_3}\Lambda^K\right]
\textstyle{\int_{S_4}}\Lambda^L
\right]
\nonumber \\
& \qquad
=
(-4\mathrm{p} \epsilon^{IJKL})
\left[\frac{-2\pi i}{\mathrm{K}}\right]^3
%I((S_1\sharp V_2 + S_2\sharp V_1)\sharp S_3, S_4),
I(\partial (V_1\sharp V_2)\sharp S_3, S_4),
\end{align}
where we noted
$d (\delta(V_1)\wedge\delta(V_2)) = \delta(S_1)\wedge  \delta(V_2) + \delta(V_1)\wedge \delta(S_2)$.
The four-loop braiding phase is encoded in the following product of Wilson operators
\begin{align}
&
\left[(W^2W^1)^{\dag}W^1W^2\cdot W^3\cdot (W^1W^2)^{\dag}W^2W^1
\cdot W^{3\dag}\right]
\nonumber \\
&\quad
\cdot W^4\cdot\left[\cdots \right]^{\dag}\cdot W^{4\dag}
\nonumber\\
&=\exp \left(
\big[
\big[\big[i\textstyle{\int_{S_1}} \Lambda^1,i
\textstyle{\int_{S_2}} \Lambda^2\big],
i
\textstyle{\int_{S_3}} \Lambda^3\big],
i
\textstyle{\int_{S_4}} \Lambda^4\big] \right).
\end{align}

\subsection{The Wilson operator algebra on $T^3$}

It is also instructive to construct the Wilson operator algebra on 
a closed spatial manifold with non-trivial topology, e.g., $\Sigma=T^3$. 
We will work in the setting identical to the previous section,
and quantize the theory on $\Sigma=T^3$.
As before, we expand $a^I$ and $b^I$ by using the Hodge decomposition
as
$
a^I = \cdots + \alpha^I_l \omega_l,
$
and
$
b^I= \cdots + \beta^I_p \eta_p
$,
where
$\alpha^I_l$ and $\beta^I_l$ are the zero modes.
Also, as before, we consider
Wilson operators associated to
the generators $\{L^m\}$ and $\{S^m\}$
of the first and second homology groups.
For $L^i$, we consider
the Wilson loop operators
\begin{align}
\hat{A}^I_i := \exp i \int_{L^i} \hat{a}^I = \exp i \hat{\alpha}^I_i
\end{align}
As for $S^m$'s,
we consider
Wilson surface operators
\begin{align}
	W^1_i
	&:=
	\exp i
	\left(
	\int_{S^i}
	b^1 -
	\mathrm{p}
	\int_{\Sigma}
	a^2\wedge a^{3} \wedge a^4
	\right)
	\quad
	\mbox{etc}.
\end{align}
The cubic term can be written as,
assuming $\Sigma$ is formal,
$
	\int_{\Sigma} a^I\wedge a^J \wedge a^K
	=
	\alpha^I_i \alpha^J_j \alpha^K_k
	\int_{\Sigma}
	\omega_i \wedge \omega_j \wedge \omega_k
%	\nonumber \\
%	&=
%	\alpha^I_i \alpha^J_j \alpha^K_k
%	\epsilon_{jkp}
%	\int_{\Sigma} 	\omega_i \wedge \eta_p
=
\epsilon_{ijk}
	\alpha^I_i \alpha^J_j \alpha^K_k.
$
Hence, the Wilson surface operators associated to $S^i$ are
\begin{align}
&
\quad
\hat{W}^I_i = \exp i \hat{\Lambda}^I_i
\nonumber \\
&
\mbox{where}
\quad
	\Lambda^I_i = \beta^I_i - \mathrm{p} \epsilon^{IJKL}\epsilon_{ijk} \alpha^J_i \alpha^K_j \alpha^L_k
%	\nonumber \\
%	\Lambda^2_i &= \beta^2_i + \mathrm{q} \epsilon_{ijk} \alpha^1_i \alpha^3_j \alpha^4_k
%	\nonumber \\
%	\Lambda^3_i &= \beta^3_i - \mathrm{q} \epsilon_{ijk} \alpha^1_i \alpha^2_j \alpha^4_k
%	\nonumber \\
%	\Lambda^4_i &= \beta^4_i + \mathrm{q} \epsilon_{ijk} \alpha^1_i \alpha^2_j \alpha^3_k
\end{align}
The Wilson operator algebra can be computed as
\begin{align}
&
	(\hat{W}^2_2 \hat{W}^1_1)^{\dag}
	\hat{W}^1_1 \hat{W}^2_2
	\nonumber \\
&	\qquad
	=
	\exp ( [i\Lambda^1_1, i\Lambda^2_2]),
	\nonumber \\
&
	(\hat{W}^2_2 \hat{W}^1_1)^{\dag}  \hat{W}^1_1 \hat{W}^2_2
	\cdot
	\hat{W}^3_3
	\cdot
	(\hat{W}^1_1 \hat{W}^2_2)^{\dag}  \hat{W}^2_2 \hat{W}^1_1
	\cdot
	\hat{W}^{3\dag}_3
%	\nonumber \\
%	&=
%	\exp\left(- [\Lambda^1_1,\Lambda^2_2] \right)
%	B^3_3
%	\exp\left(- [\Lambda^1_1,\Lambda^2_2]^{\dag}  \right)
%	B^{3\dag}_3
	\nonumber \\
	&\qquad
	=
	\exp (
	[ [i\Lambda^1_1,i\Lambda^2_2],i \Lambda^{3}_3  ]),
	\nonumber \\
&
\left[
	(\hat{W}^2_2 \hat{W}^1_1)^{\dag}  \hat{W}^1_1 \hat{W}^2_2
	\cdot
	\hat{W}^3_3
	\cdot
	(\hat{W}^1_1 \hat{W}^2_2)^{\dag}  \hat{W}^2_2 \hat{W}^1_1
	\cdot
	\hat{W}^{3\dag}_3
	\right]
	\nonumber \\
	&\quad
	\times
	\hat{W}^4_1
	\times
	\left[ \cdots \right]^{\dag}
	\times
	\hat{W}^{4\dag}_1
	\nonumber \\
	&\qquad =
	\exp( [[[ i\Lambda^1_1, i\Lambda^2_2], i\Lambda^3_3],i\Lambda^4_4] ),
	\label{four-loop Wilson algebra}
\end{align}
where the repeated commutators are given by 
\begin{align}
&
	[\Lambda^1_1, \Lambda^2_2] =
	\frac{2 \pi i \mathrm{p} }{\mathrm{K}}
	\left(
-\alpha^3_1 \alpha^4_3
+\alpha^3_2 \alpha^4_3
+\alpha^3_3 \alpha^4_1
-\alpha^3_3 \alpha^4_2
	\right),
\nonumber\\
	&[ [\Lambda^1_1,\Lambda^2_2] , \Lambda^3_3 ]
	=
	\frac{4 \pi^2 \mathrm{p}}{\mathrm{K}^2}
	(\alpha^4_1-\alpha^4_2),
	\nonumber \\
	&[[ [\Lambda^1_1,\Lambda^2_2] , \Lambda^3_3 ],\Lambda^4_4]
	=
	\frac{8\pi^3 i \mathrm{p} }{\mathrm{K}^3}.
\end{align}
The last equation in Eq.\ \eqref{four-loop Wilson algebra} with the quadruple commutator
is related to the four-loop braiding statistical process.

\section{Condensation picture}
\label{Condensation picture}

We have so far discussed the coupled $BF$ theories realizing
three-loop or four-loop braiding statistics in isolation from physical contexts.
In this section,
we try to develop physical pictures of the topological field theories discussed above.

\subsection{The $BF$ theory}

Let us start with the condensation picture of the single copy of the ordinary $BF$ theory:
\begin{align}
	S =
	\frac{i\mathrm{K}}{2\pi}
	\int b\wedge da.
	\label{pure BF Euclidean}
\end{align}
(In this section, we will work with the Euclidean action.)
The $BF$ theory can be thought of as describing
the zero correlation length limit of a gapped (topologically ordered) system,
which may arise as a result of some sort of condensation.
\cite{HanssonOganesyanSondhi2004, Balachandran1993, Chan2013, Chan2015}
There are two complimentary pictures that describe the condensation,
which are dual to each other.
In the following, we will develops these pictures by using the
duality transformations.
(We will use the equations of motion and integration over fields
for convenience, but will treat the compactification conditions on the fields 
somewhat loosely. If necessary, the compactification conditions can be treated 
rigorously by using the generalized Poisson identity.
See Ref.\ \onlinecite{Chan2015} and references therein.)

To discuss the first picture,
let us take the equation of motion $\delta S/\delta b=0$ of the $BF$ theory, which sets $da=0$.
This suggests the Meissner effect and hence the Higgs phase.
An convenient action,
in which this picture is manifest,
can be derived by integrating over $b$.
It is convenient to perturb the $BF$ theory
to go away from the strict topological limit by adding
\begin{align}
	\frac{1}{2\lambda} db \wedge\star db +
	\frac{1}{g^2}
	da\wedge \star da +i \frac{\Theta}{8\pi^2}  da\wedge da.
\label{kinetic term}
\end{align}
Here, the second and third terms are the Maxwell and axion terms for $a$, respectively,
and the first term is a two-form analogue of the Maxwell term for $b$.
The integration over $b$ can be done by making use of the equation of motion derived by taking
the functional derivative $\delta/\delta b$ of the perturbed $BF$ theory, and plug the solution
back into the action. The equation of motion can be solved as
\begin{align}
	db = -\frac{i\lambda}{2\pi} \star (d\theta+\mathrm{K}a),
\end{align}
where the scalar field $\theta$ arises as an ambiguity when integrating the equation of motion to express $b$ in terms of $a$.
Formally, the above manipulation is equivalent to dualizing the two form $b$
to the zero-form $\theta$.
The resulting effective Lagrangian is
\begin{align}
	\mathcal{L} = \frac{\lambda}{8\pi^2} (d\theta+\mathrm{K}a)^2
	+
	\frac{1}{g^2} da\wedge \star da +i \frac{\Theta}{8\pi^2}  da\wedge da.
	\label{Abelian Higgs}
\end{align}
This is nothing but the Abelian Higgs model.
\cite{HanssonOganesyanSondhi2004,Balachandran1993,Fradkin:1978dv}

Alternatively,
taking the equation of motion $\delta S/\delta a=0$ of the $BF$ theory sets $db=0$.
This suggests a two-form analogue of the Meissner effect,
which can be interpreted as arising from the condensation of monopoles in
the dual gauge field $v$ of $a$.
As before, we can integrate over $a$ in the presence of the kinetic term  \eqref{kinetic term}.
Solving the equation of motion $\delta S/\delta a=0$, $da$ can be expressed
in terms of $b$ as
\begin{align}
da =\tilde{\tau}_1 (\mathrm{K}b+dv) +i \tilde{\tau}_2\star (\mathrm{K}b+dv).
\label{sol da}
\end{align}
Here, $\tilde{\tau}_1$ and $\tilde{\tau}_2$
are the dual coupling constants
and related to the original coupling constants as
\begin{align}
&
\tilde{\tau}_1 = -\frac{\tau_1}{\tau^2_1+\tau^2_2},
\quad
\tilde{\tau}_2 = \frac{\tau_2}{\tau^2_1+\tau^2_2},
\nonumber \\
&
\tau_1 = \frac{\Theta}{2\pi},
\quad
\tau_2 = \frac{4\pi}{g^2}.
\end{align}
The one form $v$ in \eqref{sol da} arises formally as an ambiguity in solving $da$ in terms $b$.
Plugging the solution back to the action,
we obtain the effective Lagrangian for $b$ and $v$ as
\begin{align}
\mathcal{L}
&=
\frac{\tilde{\tau}_2}{4\pi}
(\mathrm{K}b + dv)\wedge \star (\mathrm{K}b+dv)
\nonumber \\
&\quad
+\frac{i \tilde{\tau}_1}{4 \pi}
(\mathrm{K}b+dv)
\wedge
(\mathrm{K}b+dv)
+\frac{1}{2\lambda}
db\wedge \star db.
\label{211}
\end{align}
This is the Julia-Toulouse-Quevedo-Trugenberger effective action that describes
the condensation of monopoles of the dual gauge field $v$.
\cite{Julia1979, Quevedo1997}

It is also instructive to have a comparison with a slightly more microscopic model, which can realize the situation described above.
For example, let us consider the Cardy-Rabinovic model
\cite{Cardy:1981qy}
\begin{align}
 Z= \mathrm{Tr}_{a, n, s}
\prod_{r} \delta[ \partial_{\mu} n_{\mu}(r) ]
 \exp(-S),
\end{align}
where
$a_{\mu}$ ($\mu=1,\ldots, 4$) is a compact $U(1)$ gauge field (an angular variable) defined on the links of the hypercubic lattice,
and $n_{\mu}$ and $s_{\mu\nu}$ are integer-valued fields defined on links and plaquettes, respectively.
%$\phi_{\mu}$ represents an Abelian gauge field.
The integer-valued two-form gauge field $s_{\mu\nu}$ amounts to allowing multivalued configurations of the gauge field.
The sum on $s_{\mu\nu}$ corresponds to a sum over topologically non-trivial configurations
with magnetic monopoles.\cite{Banks-1977} In fact, the monopole current is given explicitly by
$
m_{\mu} = ({1}/{2}) \epsilon_{\mu\nu\lambda\sigma} \partial_{\nu} s_{\lambda\sigma},
$
where $\partial_{\mu}$ is the lattice difference operator in the $\mu$-direction.
%so that the final term couples $m_{\mu}$ directly to the gauge field.
On the other hand, we interpret $n_{\mu}$ as the electric current of a charge field.
The discrete delta function $\delta[ \partial_{\mu} n_{\mu}(r) ]$ enforces current conservation.
The Boltzmann weight is given by
\begin{align}
 S
&=
 -i\mathrm{K}\sum_L n_{\mu} a_{\mu}
+ \frac{1}{2g^2} \sum_P \Gamma_{\mu\nu} \Gamma_{\mu\nu}
\nonumber \\
&\quad
 - \frac{i\mathrm{K}\theta }{32\pi^2}\sum_{r,r'}
f(r-r')
\epsilon_{\mu\nu\lambda\sigma}
\Gamma_{\mu\nu}(r)
\Gamma_{\lambda\rho}(r'),
\end{align}
where
$\Gamma_{\mu\nu} = \partial_{\mu} a_{\nu} - \partial_{\nu} a_{\mu} - 2\pi s_{\mu\nu}$
is the field strength.
The second and third terms are the Maxwell and axion terms, respectively.
(The precise nature of the smearing function $f(r-r')$ is not important here.)
%The first term in the action is a Villain-type kinetic term for the gauge field.
The sum over $n_{\mu}$ has the effect of constraining
$a_{\mu}$ to take its values restricted to  the abelian cyclic group $\mathbb{Z}_{\mathrm{K}}$,
$a_{\mu} = (2\pi/\mathrm{K}) k_{\mu}$.
%This sum itself is constrained by $\Delta_{\mu} n_{\mu}=0$, since $\phi_{\mu}$ can only couple to a conserved current.
Because the sum over $n_{\mu}$ is constrained, we can always add any total divergence to $a_{\mu}$.
Thus, the restriction to $a_{\mu}= (2\pi/\mathrm{K})k_{\mu}$ represents a partial fixing of the gauge.
%Finally, $f$ is a smearing function that interpolates between plaquette and its nearest perpendicular neighbours, which we shall generally ignore below.

For the Cardy-Rabinovic model, in the deconfined phase (charge condensation),
there are $2\pi/\mathrm{K}$ flux and the braiding with the charge leads to the fractional statistics.
The effective theory is described by the $BF$ theory.

\subsection{The three-loop braiding theories}

For the three-loop braiding theories
(either with two flavors \eqref{action 3loop} or three flavors \eqref{cubic bf three flavor}),
we can repeat the duality transformation,
which we carried out for the ordinary $BF$ theory \eqref{pure BF Euclidean}
to obtain the Abelian-Higgs model \eqref{Abelian Higgs}.
Dualizing the two-form gauge fields $b^{I}$ to scalars $\phi^I$,
we obtain an analogue of the Abelian-Higgs model
\begin{align}
	\mathcal{L}
	&=
	\sum_I
	\frac{\lambda_I}{8\pi^2}
	(a^I+d\theta^I)^2
	+
	C_{IJK}
	a^I\wedge a^J\wedge da^K
	+
	\nonumber \\
	&\quad
	+
	\sum_{I}
	\left[
	\frac{1}{g^2_I} da^I\wedge \star da^I
	+
	i \frac{\Theta_I}{8\pi^2}  da^I \wedge da^I
	\right]
	+
	\cdots
\end{align}
where we have introduced the coupling constants
$\lambda_I, g_I, \Theta_I$ for each flavor.
$C_{IJK}$ describes the cubic coupling and takes different forms
for the two- and three-flavor theories.

One can also consider an analogue of the Cardy-Rabinovic theory
for the three-loop braiding theories.
For example,
for the cubic two-flavor theory \eqref{action 3loop},
it may be considered as arising from
the following extension of the Cardy-Rabinovic theory:
\begin{align}
S&=
\frac{1}{2g^2}\sum_{P}
\Gamma^1_{\mu\nu}\Gamma^1_{\mu\nu}
+
\frac{1}{2g^2}\sum_{P} \Gamma^2_{\mu\nu}\Gamma^2_{\mu\nu}
\nonumber \\
&\qquad
+i\mathrm{K}\sum_{i,\mu}
a^1_{\mu}
\left(n^1_{\mu}-\frac{\mathrm{p}_2}{\mathrm{K}}\epsilon_{\mu\nu\lambda\rho}a^2_{\nu}\partial_{\lambda}a^1_{\rho}
\right)
\nonumber \\
&\qquad
+i\mathrm{K}\sum_{i,\mu} a^2_{\mu}
\left(n^2_{\mu}-\frac{\mathrm{p}_1}{\mathrm{K}}\epsilon_{\mu\nu\lambda\rho}a^1_{\nu}\partial_{\lambda}a^2_{\rho}
\right).
\label{CR_three}
\end{align}
The charge condensation phase of this extended
Cardy-Rabinovic theory \eqref{CR_three}
is described by the coupled $BF$ theory \eqref{action 3loop}.
%This is equivalent to adding cubic a terms $i\mathrm{p}a^1\wedge a^2\wedge da^3$.
%We note that just as one could consider different cubic terms,
%one could also consider adding different coupling terms in \eqref{CR_three}.
%This is a simple choice we make that is a minimal setting that suffices
%to illustrate the physics.

Alternatively, 
one may try to dualize the gauge fields $a^I$; 
as we have seen, in the ordinary $BF$ theory, dualizing the gauge field $a$
leads to the Julia-Toulouse-Quevedo-Trugenberger effective action
\eqref{211},
and allows us to describe the charge condensation phases as the monopole condensation
phase for the dual gauge field $v$.
Due to the cubic coupling, dualizing $a^I$ appears to be rather complicated.
The electromagnetic duality exchanges the field strength $da$ and its dual $dv$,
but this does not necessarily mean it works at the level of the connection and exchanges $a$ and $v$.
In the coupled $BF$ theories, the action is not written entirely in terms of the field strength $da^I$,
but the connections $a^I$ appear directly.

While it seems not possible to dualize all $a^I$,
we can nevertheless dualize some of $a^I$.
For example, let us consider the three-flavor theory with cubic coupling
defined in \eqref{cubic bf three flavor}.
The action is written in terms of the field strength $da^3$,
and hence one can dualize $a^3$.
As for the first and second flavors, one can dualize $b^{I=1,2}$.
The resulting action is
\begin{align}
\mathcal{L}
&=
\frac{\tilde{\tau}_2}{4\pi}
(\mathrm{K}\Lambda^3 + dv^3)\wedge \star (\mathrm{K}\Lambda^3+dv^3)
\nonumber \\
&\quad
+\frac{i \tilde{\tau}_1}{4 \pi}
(\mathrm{K}\Lambda^3+dv^3)
\wedge
(\mathrm{K}\Lambda^3+dv^3)
+\frac{1}{2\lambda_3}
db^3\wedge \star db^3
\nonumber \\
&\quad
+\sum_{I=1,2}
\Big[
\frac{\lambda_I}{8\pi^2}
(a^I +d\theta^I)^2
\nonumber \\
&\qquad \quad
+
	\frac{1}{g^2_I}
	da^I\wedge \star da^I +i \frac{\Theta_I}{8\pi^2}  da^I\wedge da^I
	\Big]
\end{align}
where
\begin{align}
\Lambda^3 = b^3 +\frac{2\pi \mathrm{p}}{\mathrm{K}} a^1 \wedge a^2.
\end{align}
Thus, after the dualization, the cubic coupling $a^1\wedge a^2 \wedge da^3$
disappears, but the magnetic condensation for the dual gauge field $v^3$
is "dressed" by $a^1$ and $a^2$.

The duality transformations can be also applied to the four-loop braiding theory, 
where the magnetic monopoles for the dual gauge field ($v^4$, say) 
are dressed by the Borromean ring formed by $a^1$, $a^2$ and $a^3$. 
Similar physical picture has been applied in constructing the wave functions for symmetry protected topological (SPT) phases,
which can be realized by proliferating domain walls decorated with an SPT phase in one lower dimension.
\cite{Xie_Chen2014}
In this respect, our models here are actually the gauged version of SPT phases.

\section{Conclusion and remarks}
\label{Conclusion and remarks}

In conclusion,
we canonically quantize the multi-flavor $BF$ theories with cubic and quartic coupling.
We study the algebra of Wilson operators to understand the three-loop and four-loop braiding processes.
Using these Wilson operators, we also construct the multiplet of ground states of
the three-loop braiding field theory on $T^3$,
and calculate the $\mathcal{S}$ and $\mathcal{T}$ matrices, which encode the fractional braiding and spin statistics.
We also discuss the topological field theory as the condensation of composite particles from some parent $U(1)$ gauge theory.

We close with a few comments on open issues.

--
In $(3+1)$d, apart from the particle-loop braiding described by the ordinary $BF$ theory,
there can be more exotic braiding, including three-loop braiding and four-loop braiding process.
In this paper, we study the topological field theory describing the three-loop braiding and four-loop braiding process.
By checking the equation of motion in Eqs.\ \eqref{eom_cubic} and \eqref{eom_quartic},
these multiple-loop braiding process can all be understood as an effective particle braiding around the loop excitation.
This effective particle can be a Hopf linking configuration, Borromean ring configuration or even more complicated knot configuration.
\begin{align}
\frac{\mathrm{K}}{2\pi}\int_{\Sigma} db^I=\int_{\Sigma}
J_{qp}^I +\mbox{``Knot configuration''}
\end{align}
It would be interesting to study more complicated knot-loop braiding process in the future.

-- In this paper, we mostly limit ourselves to $T^3$ as our spatial manifold, which is formal.
It would be interesting to study more general cases in which the coupled $BF$ theories are considered on the spacetime or spatial manifolds
which are not formal.
The coupled $BF$ theories may be able to detect topological aspects (topological invariants) of these manifolds,
which cannot be captured by the ordinary $BF$ theory.

-- We have carried out constructions of the multiplet of ground states on $T^3$ and calculated
the modular $\mathcal{S}$ and $\mathcal{T}$ matrices,
by using the basis of minimal entropy states for the ground state multiplet.
Alternatively,
the $\mathcal{S}$ and $\mathcal{T}$ matrices may be calculated by first constructing
ground states for generic (holomorphic) polarization in geometric quantization.
The action of the mapping class group of $T^3$, $SL(3,\mathbb{Z})$, on the ground state multiplet
can then be calculated by adiabatically changing polarization.
We have so far constructed ground states only for the Hodge polarization.
(See Appendix \ref{Ground state wave functionals by geometric quantization} for the definition and more details.)
Construction of the ground states for more generic polarization is left as a future problem.

%-- Finally, connection to quantum gravity (?)

\textit{Note added}: 
Upon completion of this manuscript, we became aware of a recent work by J. Wang, et al.
\cite{wang2016quantum}, which also discusses, among others, the four-loop braiding process and the connection of the quartic theory.

\acknowledgements

We thank Michael Levin for useful discussion 
and  Peng Ye for useful discussions
and for bringing Ref.\ \onlinecite{Xie_Chen2014} to our attention.
%This work is supported  by  the  NSF  under  Grant  No.  DMR-1455296, and Alfred P. Sloan foundation.
This work was supported in part by the National Science Foundation grant DMR-1408713 (XC)
and DMR-1455296 (AT and SR) at the University of Illinois,
and by Alfred P. Sloan foundation.

\appendix

\section{Ground state wave functionals by geometric quantization}
\label{Ground state wave functionals by geometric quantization}

In this section, we will construct (ground state) wave functions (functionals)
of the coupled $BF$ theories.
The ground state wave functionals of topological quantum field theories
such as the (2+1)-dimensional Chern-Simons theories
and $BF$ theories can be constructed by using the method of geometric quantization.
\cite{Bos:1989kn, BergeronSemenoffSzabo1995, Nairbook}

In geometric quantization, one endows
the phase space with a complex line bundle $E$
with curvature $\Omega$ (the symplectic two-form)
and connection $\mathcal{A}$ (the symplectic connection)
such that $\Omega$ is expressed as $\Omega=d\mathcal{A}$ (at least locally).
Sections of this line bundle form the pre-quantum Hilbert space
with element $\Psi$.
To obtain the ``physical'' Hilbert space 
which implements unitarity and irreducibility on the Poisson bracket,
one further needs to impose a constraint on $\Psi$.
This procedure is called choosing polarization.
For more details of geometric quantization,
see, Ref.\ \onlinecite{Nairbook}, for example.

\subsubsection{two-flavor v.s. three-flavor theories}
In the following, we will construct the ground state wave functions of the coupled $BF$ theories
on $\Sigma=T^3$.
We will focus on the quadratic avatar of the three flavors of $BF$ theories coupled by a cubic term.
\begin{align}
S&=
\int_{\mathcal M}
\Big\{
\frac{\mathrm{K}}{2\pi}\delta_{IJ}b^{I}\wedge da^{J}
-\delta_{IJ}a^{I}\wedge J_{qp}^{J} 
\nonumber \\
&\quad 
\quad 
-\left[b^{1}+\frac{\mathrm{q}_1}{2\pi}a^{2}\wedge a^{3}\right]\wedge J_{qv}^{1} 
\nonumber \\
&\quad \quad 
-\left[b^{2}+\frac{\mathrm{q}_2}{2\pi}a^{3}\wedge a^{1}\right]\wedge J_{qv}^{2}
\nonumber \\
& \quad \quad 
-\left[b^{3}+\frac{\mathrm{q}_3}{2\pi}a^{1}\wedge a^{2}\right]\wedge J_{qv}^{3}
\Big\}.
\end{align}
Furthermore, we will focus on the zero mode sector.
(The wave functions of the ``oscillator'' part of the theory is identical to those in the ordinary $BF$ theory,
and can be constructed by following, e.g., Ref.\ \onlinecite{BergeronSemenoffSzabo1995}. 

Working with the three-flavor theory has a technical advantage than the two-flavor theory.
To explain the advantage, we split the construction of the ground state wave functions in the following two steps:

(i) One first identifies the symplectic structure of the zero mode phase space.
Then, following the generic procedure of the geometric quantization,
one chooses the polarization (i.e., the choice of variables to use to write down wave functions).
One can then identify the generic structure of the wave functions, inner product, etc.
We call the set of wave functions obtained this way the ``large'' Hilbert space.

(ii)
The ``large'' Hilbert space is not yet of our physical relevance, since
they are not invariant under large gauge transformations.
To further write down ground state wave functions explicitly,
we need to demand the large gauge invariance (the Gauss law constraint).
(Since systems of our interest are topological and there is no Hamiltonian.
The large gauge invariance is the only guidance to construct physical ground state wave functions.)
We demand the set of the wave functions are gauge-singlet
(or in fact one can relax this condition a little bit;
one may demand the wave functions to form a projective representation
of the algebra of the gauge transformations.
Such ``generalized'' gauge invariance is in particular relevant when the level $\mathrm{K}$ is
a rational number $\mathrm{K}=\mathrm{k}_1/\mathrm{k}_2$. Here, we will focus on the
simplest case when $\mathrm{K}=\mbox{integer}$ or $\mathrm{k}_2=1$).

For the two-flavor theory, the main difficulty is that,
the large gauge transformations cannot be represented as a unitary operator within the ``large'' Hilbert space.
This can be seen from the fact that the set of commutators are not preserved by the large gauge transformations.
(See Sec.\ \ref{Large gauge invariance two flavors}.)
In other words, the symplectic two-form is not preserved under the large gauge transformation.
This should be contrasted to the case of the (2+1)-dimensional Chern-Simons theory and the ordinary $BF$ theories in (3+1) dimensions.
That the large gauge transformations cannot be represented as unitary operators within the large Hilbert space
does not mean that it is impossible to construct the ``small'' or restricted Hilbert space which is gauge invariant.
Nevertheless, this difficulty adds some complication in constructing the ground state wave functions.

For the three-flavor theory, there is no such difficulty;
the symplectic two-form is manifestly large gauge invariant;
the technical reason why we will work with the three-flavor theory in this section.

\subsubsection{choice of polarization}

There is another complication in quantizing and constructing wave functions in coupled $BF$ theories,
which is associated to the choice of polarization.
In the (2+1)-dimensional Chern-Simons theories and $BF$ theories,
it is convenient to choose a generic holomorphic polarization.
In the case of the Chern-Simons theory, this is convenient when making a contact with
(1+1)-dimensional conformal field theories.
In the coupled $BF$ theory, however, we will focus on a specific polarization,
the ``Hodge'' polarization
following the terminology in Ref.\ \onlinecite{Dunne:1989it}.
In this polarization, we construct wave functions in terms of the zero modes $\alpha^I_i$.
One reason for this is that we found it is somewhat technically involved
to construct the wave function by using the holomorphic polarization.
However, on the other hand,
the comparison
with wave functions constructed in Sec. \ref{Wave function in terms of Wilson operators}
can be easily made
for the wave functions in the Hodge polarization.

\subsection{Geometric quantization of the $BF$ theory}

We now move on to the construction of wave functions by geometric quantization.
We start by taking the ordinary $BF$ theory on $\mathcal{M}=T^3 \times \mathbb{R}$ as an example.
Our setting is described in Sec.\ \ref{Quantization on a closed spatial manifold}.
%We move on to the quantization in the zero mode sector.
As mentioned earlier, we will focus on the zero mode sector.
The zero modes of the $BF$ theory satisfy the Poisson bracket
\begin{align}
\left\{
\alpha_i, \beta_j
\right\}
=
\frac{2\pi}{\mathrm{K}} \delta_{ij}.
\end{align}

\subsubsection{the holomorphic polarization}

Let us first construct wave functions in the holomorphic polarization
following Ref.\ \onlinecite{BergeronSemenoffSzabo1995}.
In the holomorphic polarization,
we introduce complex coordinates
\begin{align}
\gamma_{i} &:=\alpha_{i}+\rho_{ij}\beta_{j},
\nonumber \\
\bar{\gamma}_i &:=\alpha_{i}+\bar{\rho}_{ij}\beta_{j},
\label{Holomorphic polarization}
\end{align}
where $\rho$ is an arbitrary symmetric $3\times 3$ complex-valued matrix, whose
imaginary part is negative-definite.
$\rho$ can be thought of as parametrizing
a complex structure on $H^1(\Sigma;\mathbb{R})\oplus H^2(\Sigma;\mathbb{R})$
forming the multi-dimensional complex space of the $\gamma$ variables.
\cite{BergeronSemenoffSzabo1995}
%and this then determines all of the topological degrees of freedom which remain
%in the source-free case modulo the large gauge transformations (4.20).
%In fact, as we shall see later on, the positive-definite symmetric matrix
%actually defines a metric on P, and from its definition (4.22) (and the definition (4.19)) we
%see that it incorporates the topological linking of the homology cycles of M3 .
%However,
%since the action (2.1) defines a topological field theory, all observables will be independent
%of this phase space complex structure. This is analogous to the situation in Chern-Simons
%theory [21]. From (4.24) and (4.19) we see that the quantization of the classical phase
%space of the BF system will give (projective) quantum representations of the cohomology
%groups of M3 .
The inverse transformations are
\begin{align}
\beta_{i}&=
\frac{1}{2i} R_{ij} (\gamma-\bar{\gamma})_j,
\nonumber \\
\alpha_i &=
\frac{-1}{2i} (\bar{\rho}R \gamma - \rho R\bar{\gamma})_i,
\end{align}
where we introduced the notation
\begin{align}
	(\mathrm{Im}\, \rho)^{-1}_{ij}= R_{ij}
\end{align}
The complex coordinates satisfy
the Poisson bracket
\begin{align}
	\left\{
	\gamma_i, \bar{\gamma}_j
	\right\}
%	=
%	\frac{2\pi}{\mathrm{K}}
%	(\bar{\rho}^i_j - \rho^j_i)
%	\delta^{IJ}
	=
	\frac{2\pi}{\mathrm{K}}
	(-2i \mathrm{Im}\, \rho^{i}_j).
\end{align}
The symplectic 2-form is
\begin{align}
	\Omega =
	-
	\frac{i\mathrm{K}}{4\pi}
	R_{ij}
	d\bar{\gamma}_i
	\wedge
	d\gamma_j
\end{align}
We choose the symplectic potential as
\begin{align}
	\mathcal{A}
	=
	+\frac{\mathrm{K}}{8\pi}
	(\bar{\gamma}-\gamma)_i
	R_{ij}
	(\bar{\rho}R d\gamma- \rho R d\bar{\gamma})_j,
\end{align}
which satisfies $d\mathcal{A}=\Omega$.

As a first step of constructing ground state wave functions,
we choose a particular polarization and impose
the condition:
\begin{align}
&
\left(
	\frac{\partial}{\partial \bar{\gamma}_i}
	+
	i \mathcal{A}_{\bar{\gamma}_i}
	\right)
	\Psi
	=0
	\nonumber \\
	&\Rightarrow
\left(
	\frac{\partial}{\partial \bar{\gamma}_i}
        -
	i
	\frac{\mathrm{K}}{8\pi}
	[(\bar{\gamma}-\gamma) R \rho R]_i
	\right)
	\Psi
	=0.
\end{align}
Solutions to this constraint are given by
\begin{align}
\Psi(\gamma,\bar{\gamma}) =
	\exp \left[
	-i
	\frac{\mathrm{K}}{16\pi}
	(\bar{\gamma}-\gamma) R \rho R
	(\bar{\gamma}-\gamma)
	\right]
	f(\gamma),
\end{align}
where $f$ is a function of $\gamma$ only.
The set of all wave functions of the above form constitute what we have called the ``large'' Hilbert space.

We now construct a set of ground state wave functions
by imposing the invariance under large gauge transformations
\begin{align}
	\gamma \to \gamma + 2\pi (n + \rho m).
\end{align}
In the following, we present two slightly different construction of the wave functions.

In the first construction, we note,
under the large gauge transformations,
the symplectic potential is transformed as
\begin{align}
&
	\mathcal{A}
	\to
	\mathcal{A}
	+
	d\Lambda
	\nonumber \\
	&
	\mbox{where}
	\quad
	\Lambda =
-
	\frac{\mathrm{K}i}{2}
	m
	\cdot
	(\bar{\rho}R \gamma- \rho R \bar{\gamma})
	+
	const.
\end{align}
where the constant term can depend on $m$ and $n$.
Physical wave functions, which are gauge invariant, then must satisfy
\begin{align}
	\Psi(\gamma+2\pi (n+\rho m),
	\bar{\gamma}+2\pi (n+\bar{\rho} m))
	=
	e^{ i\Lambda}
	\Psi(\gamma, \bar{\gamma})
\end{align}
%Noting
%\begin{align}
%&
%	\exp \left[
%	-i \frac{\mathrm{K}}{16\pi}
%	(\bar{\gamma}-\gamma) R \rho R
%	(\bar{\gamma}-\gamma)
%	\right]
%	\nonumber \\
%	&\rightarrow
%	\exp \left[
%	-i \frac{\mathrm{K}}{16\pi}
%	(\bar{\gamma}-\gamma )
%	R \rho R
%	(\bar{\gamma}-\gamma )
%	\right]
%	\nonumber\\
%	&\quad
%	\times
%	\exp \left[
%	-  \frac{\mathrm{K}}{2}
%	 m\cdot
%	 \rho R
%	 \cdot
%	(\bar{\gamma}-\gamma )
%	\right]
%	\times
%	\exp \left[
%	+i \pi \mathrm{K}
%	m
%	\cdot
%	 \rho
%	 \cdot
%	m
%	\right]
%\end{align}
This condition is translated into the condition on $f$:
\begin{align}
&
e^{
	+ i \mathrm{K} m\cdot \gamma
	+i \pi \mathrm{K}
	m
	\cdot
	 \rho
	 \cdot
	m
	}
	f(\gamma + 2\pi (n+m\rho) )
	=
	f(\gamma)
	\label{constraint f}
\end{align}
up to an unknown phase factor mentioned above.
The solution can be constructed by
using the Jacobi theta function:
\begin{align}
\Psi_q(\gamma)=
\Theta\left(\begin{array}{c}
	            	\frac{c+q}{\mathrm{K}}\\
	            	d
	            \end{array}
\right)
\left(
\frac{\mathrm{K}}{2\pi} \gamma
|
-\mathrm{K}\rho
\right),
\end{align}
where $c$ and $d$ are arbitrary parameters (``twisting angles''). 
Here, the Jacobi theta function is defined by 
\begin{widetext}
\begin{align}
&
	\Theta\left(\begin{array}{c}
	            	c \\
	            	d
	            \end{array}
\right)(z|\Pi)
:=
\sum_{n_{\ell} \in \mathbb{Z}^p}
\exp\left[
i\pi (n+c)^{\ell} \Pi_{\ell k} (n+c)^k
+
2\pi i (n+c)^{\ell} (z+d)_{\ell}
\right]
\end{align}
where $c^{\ell}, d_{\ell}\in [0,1]$ and $z_{\ell}\in \mathbb{C}$.
The theta function satisfies
\begin{align}
	\Theta\left(\begin{array}{c}
	            	c \\
	            	d
	            \end{array}
\right)(z_{\ell} + s_{\ell} + \Pi_{\ell k} t^k|\Pi)
&:=
\exp\left[
2\pi i c^{\ell} s_{\ell} - i \pi t^{\ell} \Pi_{\ell k} t^k - 2\pi i t^{\ell} (z + d)_{\ell}
\right]
	\Theta\left(\begin{array}{c}
	            	c \\
	            	d
	            \end{array}
\right)(z|\Pi)
\end{align}
for integers $s_l$ and $t_l$, and
\begin{align}
		\Theta\left(\begin{array}{c}
	            	c \\
	            	d
	            \end{array}
\right)(z_{\ell} + C \Pi_{\ell k} t^k|\Pi)=
\exp\left[
- i \pi C^2 t^{\ell} \Pi_{\ell k} t^k - 2\pi i C t^{\ell} (z+d)_{\ell}
\right]
	\Theta\left(\begin{array}{c}
	            	c + Ct\\
	            	d
	            \end{array}
\right)(z|\Pi)
\end{align}
for any non-integer $C\in \mathbb{R}$.
We note, in particular,
\begin{align}
&
\Theta\left(\begin{array}{c}
	            	\frac{c+q}{\mathrm{K}}\\
	            	d
	            \end{array}
\right)
\left(
\frac{\mathrm{K}}{2\pi} [\gamma+2\pi (n + \rho m) ]
|
-\mathrm{K}\rho
\right)
\nonumber \\
&=
\exp\left[
2\pi i
c \cdot n
+
i \pi m \cdot \mathrm{K}\rho \cdot m
+
i 2\pi m\cdot d
\right]
\exp (i \mathrm{K} m\cdot  \gamma)
\Theta\left(\begin{array}{c}
	            	\frac{c+q}{\mathrm{K}}\\
	            	d
	            \end{array}
\right)
\left(
\frac{\mathrm{K}}{2\pi} \gamma
|
-\mathrm{K}\rho
\right)
\end{align}
\end{widetext}

In the second construction,
we implement the large gauge transformation by using
unitary operators, which we call $U_{m,n}$.
This operator sends $\alpha\to \alpha+2\pi n$ and $\beta \to \beta+2\pi m$:
\begin{align}
U_{m,n} \alpha U^{\dag}_{m,n}=\alpha+2\pi n,
\quad
U_{m,n} \beta U^{\dag}_{m,n}=\beta+2\pi m.
\label{large gauge trsf}
\end{align}
The unitary operator can be identified, up to
a constant phase factor, as
\begin{align}
	U_{m,n} =
	\exp\left[ -i \mathrm{K} (m_i \alpha_i - n_i \beta_i)
	\right].
\end{align}
Noting
$
\bar{\gamma}_i =
-({4\pi}/{\mathrm{K}} )R^{-1}_{ij} (\partial/\partial\gamma_j),
$
the operator implementing the large gauge transformations can be written as
\begin{align}
	U_{m,n}
	&=
	\exp
	\left[
	2\pi (n + m\cdot \rho)\cdot \frac{\partial}{\partial\gamma}
	+
	\frac{\mathrm{K}}{2}
	( n+m\cdot \bar{\rho} )\cdot R \cdot \gamma
	\right]
	\nonumber \\
		&=
e^{	
	\frac{\pi \mathrm{K}}{2} (n+m\bar{\rho}) R (n+m\rho)
	+
	\frac{\mathrm{K}}{2} (n+m\bar{\rho}) R \gamma
	}
\nonumber \\
&\quad
\times
\exp\left[
2\pi (n+m\rho) \frac{\partial}{\partial\gamma}
\right]
\end{align}
The action of $U$ on wave functions is
\begin{align}
	U_{m,n}\Psi (\gamma)
	&=
	e^{
	\frac{\pi \mathrm{K}}{2} (n+m\bar{\rho}) R (n+m\rho)
	+
	\frac{\mathrm{K}}{2} (n+m\bar{\rho}) R \gamma
	}
\nonumber \\
&\quad \times
\Psi(\gamma+2\pi (n+\rho m))
\end{align}
The wave functions that solve this constraint are given by
\begin{align}
	\Psi_q(\gamma)
&=
e^{ -\frac{\mathrm{K}}{8\pi} \gamma_{i} R_{ij} \gamma_{j} }
\Theta\left(\begin{array}{c}
	            	\frac{c+q}{\mathrm{K}}\\
	            	d
	            \end{array}
\right)
\left(
\frac{\mathrm{K}}{2\pi} \gamma
|
-\mathrm{K}\rho
\right),
\end{align}
where $c$ and $d$ are arbitrary parameters (``twisting angles''). 
%Note that
%\begin{align}
%&
%e^{ \frac{\mathrm{K}}{4\pi} (\gamma+2\pi (n+\rho m))^{\ell} (\Omega^{-1})_{\ell k} (\gamma+2\pi (n+\rho m))^{k} }
%\nonumber \\
%&=
%e^{ \frac{\mathrm{K}}{4\pi} \gamma^{\ell} (\Omega^{-1})_{\ell k} \gamma^{k} }
%\nonumber \\
%&\quad
%\times
%e^{ \mathrm{K} \gamma^{\ell} (\Omega^{-1})_{\ell k} (n+\rho m)^{k} }
%e^{ \pi \mathrm{K}  (n+\rho m)^{\ell} (\Omega^{-1})_{\ell k}  (n+\rho m)^{k} }
%\end{align}
%and
%\begin{align}
%	- \frac{\mathrm{K}}{2} \gamma R^{-1} \rho m
%	&=
%	- \frac{\mathrm{K}}{2} \gamma R^{-1} \bar{\rho} m
%	- \frac{\mathrm{K}}{2} \gamma R^{-1} (\rho-\bar{\rho}) m
%	\nonumber \\
%	&=
%	-\frac{\mathrm{K}}{2} \gamma R^{-1} \bar{\rho} m
%	+
%	\mathrm{K} i m\cdot \gamma
%\end{align}
%
%
\subsubsection{the Hodge polarization}

We have so far constructed wave functions
by using the holomorphic polarization \eqref{Holomorphic polarization}.
We now try a different poloarization, which we call
the Hodge polarization, following,
Ref.\  \onlinecite{Dunne:1989it}.
In this polarization,
we attempt to write down the wave function in terms of $\alpha_i$:
$
	\Psi(\alpha)
$.
Given the canonical commutation relation
$
[\alpha_i, \beta_j] =(2\pi i/\mathrm{K}) \delta_{ij}
$,
$\beta_i$ acts on the wave functions as
$
	\beta_i = -i (2\pi/\mathrm{K}) \partial/\partial \alpha_i
$.
Demanding \eqref{large gauge trsf},
the unitary transformations that implement
large gauge transformations can be represented as
\begin{align}
	U_{m,n} &=
	\exp\left[
	-i\mathrm{K} m \cdot \alpha+i \mathrm{K} n \cdot \beta
	\right]
%	\nonumber \\
%	&=
%	\exp\left[
%	-i\mathrm{K} m \cdot \alpha
%	+2\pi  n \cdot
%	\frac{\partial}{\partial \alpha}
%	\right]
	\nonumber \\
	&=
	e^{ - \pi i \mathrm{K} m\cdot n -i\mathrm{K} m \cdot \alpha }
	\exp\left[
	2\pi   n \cdot
	\frac{\partial}{\partial \alpha}
	\right].
\end{align}
Physical wave functions can be constructing by demanding large gauge invariance:
\begin{align}
	U_{m,n} \Psi(\alpha) = e^{i\Theta_{m,n}} \Psi(\alpha)
\end{align}
where $\Theta_{m,n}$ is a constant phase, which can depend on $m$ and $n$.
I.e.,
\begin{align}
	U_{m,n}\Psi(\alpha)
	&=
	e^{  -\pi i \mathrm{K} m\cdot n
	-i \mathrm{K} m \cdot \alpha
	}
	\Psi(\alpha+2\pi n)
	\nonumber \\
	&
=
e^{i\Theta_{m,n}} \Psi(\alpha)
\end{align}
This constraint can be solve by an ansatz
\begin{align}
	\Psi(\alpha) &=
	\sum_{k\in \mathbb{Z}^3} C(k) e^{i k\cdot \alpha}.
\end{align}
From the large gauge invariance,
$C$ must satisfy the constraint
\begin{align}
	C(p+\mathrm{K}m) = e^{i\theta} C(p),
\end{align}
which can be solved by
\begin{align}
	C_q (p) = 
	\left\{
	\begin{array}{ll}
	e^{i\theta l}
	&
	\mbox{when} \quad p = q+ \mathrm{K}l \\
	0
	&
	\mbox{otherwise}
	\end{array}
\right.	
\end{align}
To summarize, the solutions are
\begin{align}
	\Psi_q(\alpha) &=
	e^{i \theta \cdot \alpha} e^{iq\cdot \alpha}
	\sum_{l}
	e^{i \mathrm{K} l \cdot (\alpha+\phi/\mathrm{K}) }
	\nonumber \\
	&=
	e^{i \theta \cdot \alpha} e^{iq\cdot \alpha}
      \frac{2\pi}{\mathrm{K}}
      \sum_m
      \delta\left( \alpha+ \frac{\phi}{\mathrm{K}} + \frac{2\pi}{\mathrm{K}} m\right).
\end{align}
The free parameter $\phi$ and $\theta$ are the twisting angle. 

The states we have constructed are eigen states of
$
B_i = \exp i \beta_i = \exp [(2\pi/\mathrm{K}) \partial/\partial \alpha_i]
$.
On the other hand, applying $A_i$ changes the label $q$
as
$A_i \Psi_q =\Psi_{q+\hat{n}_i}$, 
where $\hat{n}_i=(0,\cdots, 1, \cdots, 0)$.
%\begin{align}
%	A_i |\Psi_q\rangle = e^{i \alpha_i} \int d\alpha\Psi_q(\alpha) |\alpha\rangle
%	=
%	\int d\alpha \Psi_{q+\hat{n}_i}(\alpha)|\alpha\rangle
%	=
%	|\Psi_{q+\hat{n}_i}\rangle
%\end{align}
%where $\hat{n}_i$ is $(0,\cdots, 1, \cdots, 0)$.
%

\subsection{Three-loop braiding theory with three flavors}

We now move on to the construction of wave functions
of the quadratic three loop braiding $BF$ theory with three flavors.
The zero modes of the three-flavor theory satisfy the Poisson bracket
\begin{align}
\left\{
\alpha^I_i, \beta^J_j
\right\}
=
\frac{2\pi}{\mathrm{K}} \delta_{ij}\delta^{IJ}.
\end{align}
The symplectic form and potential ($I,J=1,2,3$) are given by
\begin{align}
	\Omega &= \frac{\mathrm{K}}{2\pi} d\beta^I_i \wedge d \alpha^I_i,
	\quad
	\mathcal{A}= \frac{\mathrm{K}}{2\pi} \beta^I_i d\alpha^I_i.
\end{align}

In the quadratic three-flavor $BF$ theory,
the fundamental Wilson surface operators are defined by taking
the exponential of
\begin{align}
	\Lambda^I_i = \beta^I_i + \mathrm{r} \epsilon^{IJK}_{ijk} \alpha^J_j \alpha^K_k,
\end{align}
as $\exp i \Lambda^I_i$, where we have introduced
$\epsilon^{IJK}_{ijk}=\epsilon^{IJK}\epsilon_{ijk}$.
Generic Wilson surface operators are given by taking products thereof.
The parameter $\mathrm{r}$ plays a role similar to $\mathrm{q}_{1,2}$ in the two-flavor theory,
%will be determined later by demanding that the dimension of the the ground state multiplet is finite:
%\begin{align}
%	\mathrm{r} = \frac{\mbox{integer}}{4\pi}.
%\end{align}

There is a set of large gauge transformations that preserve $\Lambda^I_i$:
\begin{align}
\alpha^I_i &\to \alpha^I_i + 2\pi n^I_i
\nonumber \\
\beta^I_i &\to \beta^I_i - 4\pi \mathrm{r} \epsilon^{IJK}_{ijk} n^J_j \alpha^K_k
- 4\pi^2 \mathrm{r} \epsilon^{IJK}_{ijk} n^J_j n^K_k
\label{large gauge trsf 3 flavor}
\end{align}
The symplectic form is invariant under these large gauge transformations.
%as seen from
%\begin{align}
%	d\beta^I_i \wedge d\alpha^I_i
%	\to
%	d\beta^I_i \wedge d\alpha^I_i
%	-
%	4\pi \mathrm{r} \epsilon^{IJK}_{ijk} n^J_j d\alpha^K_k \wedge d\alpha^I_i,
%\end{align}
%where the last term vanishes due to the antisymmetry of the wedge product.
Under the large gauge transformations,
the symplectic form is transformed as
\begin{align}
	\mathcal{A} &\to \mathcal{A}
	+d\Lambda
	\nonumber \\
	\Lambda &=
	+\mathrm{K} m^I_i \alpha^I_i
	-\mathrm{K}\mathrm{r}
	\epsilon^{IJK}_{ijk} n^J_j \alpha^K_k \alpha^I_i
	\nonumber \\
	&\qquad
- 2\pi \mathrm{K}\mathrm{r} \epsilon^{IJK}_{ijk} n^J_j n^K_k \alpha^I_i
+
const.
\end{align}

In the following, we will write down a set of ground state wave functions for the
quadratic three-flavor $BF$ theory.
We present two different constructions.
In the first construction, we choose to
work with $\alpha^I_i$ and $\Lambda^I_i$.
Following the previous section, we introduce a holomorphic polarization for these variables.
A merit of this construction is that the large gauge transformations act on these variables
in a simple fashion.
In the second construction, we choose to work with $\alpha^I_i$ and $\beta^I_i$,
and use the Hodge polarization.
Unlike $\Lambda^I_i$, the large gauge transformations act on $\beta^I_i$ non-trivially.

\subsubsection{using $\Lambda$ as a variable}

Following the holomorphic polarization of the ordinary $BF$ theory
\eqref{Holomorphic polarization},
we introduce
\begin{align}
	\gamma^I_i &= \alpha^I_i + \rho_{ij}\Lambda^I_j,
	\nonumber \\
	\bar{\gamma}^I_i &= \alpha^I_i + \bar{\rho}_{ij}\Lambda^I_j.
\end{align}
%The inverse transformation is
%\begin{align}
%	\Lambda^I_i &= \frac{1}{2i} (R\gamma^I-R\bar{\gamma}^I)_i,
%	\nonumber \\
%	\alpha^I_i &= \frac{-1}{2i} (\bar{\rho}R\gamma^I-\rho R\bar{\gamma}^I)_i.
%\end{align}
%In terms of the complex coordinates,
%\begin{align}
%	\mathcal{A}
%	&=
%	\frac{\mathrm{K}}{8\pi} (R\gamma-R\bar{\gamma})^I_i (\bar{\rho} R d\gamma- \rho R d\bar{\gamma})^I_i
%	\nonumber \\
%	&
%-
%	\frac{\mathrm{K}\mathrm{r}}{16\pi i}
%	\epsilon_{IJK}\epsilon_{ijk}
%	\nonumber \\
%	&\quad
%	\times
%	(\bar{\rho} R \gamma- \rho R\bar{\gamma})^J_j
%	(\bar{\rho} R \gamma- \rho R\bar{\gamma})^K_k
%	(\bar{\rho} R d\gamma- \rho Rd\bar{\gamma})^I_i
%\end{align}
The wave functions can be constructed by demanding
\begin{align}
	\left(\frac{\partial}{\partial \bar{\gamma}^I_i} + i \mathcal{A}_{\bar{\gamma}^I_i}
	\right)
	\Psi(\gamma,\bar{\gamma})=0.
\end{align}
The solutions to this constraint are given by
\begin{align}
	\Psi(\gamma,\bar{\gamma})
	&=
	\exp
	\Bigg[
	  \frac{i\mathrm{K}}{16\pi}
	 \Lambda^I_i
	 \rho_{ij}
	 \Lambda^I_j
%	 (R\gamma-R\bar{\gamma})^I_i \rho_{ij} (R\gamma-R\bar{\gamma})^J_j
-
	\frac{i\mathrm{K}\mathrm{r}}{6\pi} \epsilon^{IJK}_{ijk}
	\alpha^I_i
	\alpha^J_j
	\alpha^K_k
%	(\bar{\rho} R \gamma- \rho R\bar{\gamma})^I_i
%	(\bar{\rho} R \gamma- \rho R\bar{\gamma})^J_j
%	(\bar{\rho} R \gamma- \rho R\bar{\gamma})^K_k
	\Bigg]
	f(\gamma),
\end{align}
where $f(\gamma)$ is a function of $\gamma$ only.
%
%\begin{align}
%	\Psi
%	&=
%	\exp
%	\Big[
%	- i \frac{K}{16\pi} (R\gamma-R\bar{\gamma})^I_i \rho_{ij} (R\gamma-R\bar{\gamma})^J_j
%	\nonumber\\
%	&\quad
%	-
%	\frac{Kq}{48\pi} \epsilon_{IJK} \epsilon_{ijk}
%	\nonumber \\
%	&\quad
%	\times
%	(\bar{\rho} R \gamma- \rho R\bar{\gamma})^I_i
%	(\bar{\rho} R \gamma- \rho R\bar{\gamma})^J_j
%	(\bar{\rho} R \gamma- \rho R\bar{\gamma})^K_k
%	\Big]
%	\nonumber \\
%	&\quad
%	\times
%	f(\gamma)
%\end{align}
%

We now impose the large gauge invariance.
Up to a constant phase factor,
$f$ must transform as
\begin{align}
&
e^{
	+ i \mathrm{K} m^I \cdot \gamma^I
	+i \pi \mathrm{K}
	m^I
	\cdot
	 \rho
	 \cdot
	m^I
-
	\frac{4 i\pi^2 \mathrm{K} \mathrm{r}}{3}
	\epsilon_{IJK}
	(n^I \cdot n^J\times n^K)
}
\nonumber \\
&\quad 
\times 
	f(\gamma + 2\pi (n+\rho\cdot m) )
	\nonumber \\
	&\quad
	=
	f(\gamma)
\end{align}
Up to the phase factor, this constraint is the same
as the one in the ordinary $BF$ theory.
Hence, the solutions to the gauge constraint
are given in terms of the theta function.

\subsubsection{the Hodge polarization}

We now attempt to construct the wave functions by using the Hodge polarization,
in which
the wave functions are constructed as a function of $\alpha^I_i$.
On these wave functions, $\beta^I_i$ acts as
$
	\beta^I_i = -i ({2\pi}/{\mathrm{K}}) \partial/\partial \alpha^I_i
$.
We first look for unitary operators $U_{m,n}$ that implement
the large gauge transformations \eqref{large gauge trsf 3 flavor}.
%\begin{align}
%&
%U_{m,n} \alpha^{I}_i U^{\dag}_{m,n} =
%\alpha^I_i + 2\pi n^I_i,
%\nonumber \\
%&
%U_{m,n} \beta^{I}_i U^{\dag}_{m,n} =
%\beta^I_i + 2\pi m^I_i
%\nonumber \\
%&\qquad
%- 4\pi \mathrm{r} \epsilon_{ijk}^{IJK} n^J_j \alpha^{K}_k
%- 4\pi^2 \mathrm{r} \epsilon^{IJK}_{ijk} n^J_j n^K_k.
%\end{align}
Up to a phase factor, the unitary operators $U_{m,n}$ are identified as
\begin{align}
	U_{m,n}
	&=
	\exp\Big[
	-i\mathrm{K} m^I \cdot \alpha^I
	+i \mathrm{K} n^I \cdot \beta^I
	\nonumber \\
	&\quad
	+ i \mathrm{K} \mathrm{r}
	\epsilon^{IJK}
	n^I \cdot \alpha^J \times \alpha^K
	\Big]. 
%	\nonumber \\
%	&=
%	\exp\Big[
%	-i\mathrm{K} m^I \cdot \alpha^I
%	 + i \mathrm{K} \mathrm{r}
%	 \epsilon^{IJK}
%	 n^I \cdot
%	 \alpha^J\times  \alpha^K
%	 \nonumber \\
%	 &\qquad
%	+2\pi  n^I \cdot
%	\frac{\partial}{\partial \alpha^I}
%	\Big]
\end{align}
$U_{m,n}$ can also be written as
\begin{align}
U_{m,n}
&=
e^{i \phi_{n,m}} e^A e^B
\end{align}
where
\begin{align}
&
A =
	-i\mathrm{K} m^I \cdot \alpha^I
	+ 2\pi i \mathrm{K} \mathrm{r} \epsilon^{IJK} n^I \cdot n^J\times \alpha^K
	\nonumber \\
	&\qquad
	 + i\mathrm{K}\mathrm{r}  \epsilon^{IJK} n^I \cdot \alpha^J\times  \alpha^K,
	 \nonumber \\
&B=
	+2\pi  n^I \cdot \frac{\partial}{\partial \alpha^I},
	\nonumber \\
	&
	\phi_{n,m}=
	-i \pi \mathrm{K} m^I \cdot n^I
+\frac{4\pi^2 i}{3} \mathrm{K}\mathrm{r}  \epsilon^{IJK} n^I \cdot (n^J \times n^K)
\end{align}

The physical wave functions are constrained by the large gauge invariance and must satisfy:
$
	U_{m,n} \Psi(\alpha) = e^{i\Theta_{m,n}} \Psi(\alpha).
$
This large gauge constraint can be solved by the ansatz
\begin{align}
	\Psi(\alpha) &= e^{ -\frac{i \mathrm{K}\mathrm{r}}{6\pi} \epsilon^{IJK} \alpha^I \cdot (\alpha^J\times \alpha^K) }
	\prod_{I=1}^3
	\sum_{k_I} C_I(k_I) e^{i k^I \cdot \alpha^I}
\end{align}
Observe that this wave function can be also written as
\begin{align}
	\Psi(\alpha) &=
	\prod_{I=1}^3
	\sum_{k_I} C_I(k_I) e^{i (k^I - \frac{\mathrm{K}\mathrm{r}}{18\pi}\epsilon^{IJK}  \alpha^J\times \alpha^K) \cdot \alpha^I}
%	\nonumber \\
%	&\quad
%	=
%	\sum_{k} C_1(k) e^{i (k-\frac{Kq}{3\pi} \alpha^2\times \alpha^3) \cdot \alpha^1}
%	\nonumber \\
%	&\quad
%	\times
%	\sum_{k} C_2(k) e^{i (k-\frac{Kq}{3\pi} \alpha^3\times \alpha^1) \cdot \alpha^2}
%	\sum_{k} C_3(k) e^{i (k-\frac{Kq}{3\pi} \alpha^2\times \alpha^1) \cdot \alpha^3}
	\nonumber \\
	&=
	\sum_{k} C_1(k^1) e^{i (k^1-\frac{\mathrm{K}\mathrm{r}}{\pi} \alpha^2\times \alpha^3) \cdot \alpha^1}
 \prod_{I=1}^2
	\sum_{k_I} C_I(k_I) e^{i k^I \cdot \alpha^I}.
\end{align}
The large gauge invariance constrains $C_I$ to satisfy 
\begin{align}
	C_I(p+\mathrm{K}m)
	 = e^{i\theta} C_I(p),
\end{align}
which can be solved by the same ansatz as in the ordinary $BF$ theory,
\begin{align}
	C_{Iq} (p) = e^{i\theta l}
	\quad
	\mbox{when}
	\quad
	p = q+ \mathrm{K}l
\end{align}

\bibliography{reference}

\end{document}